\documentclass[lettersize,journal]{IEEEtran}

\usepackage{soul}
\usepackage{amsmath,epsfig,amssymb,verbatim,amsopn,cite,multirow}
\usepackage{amsthm}
\usepackage{balance}
\usepackage{multirow}
\usepackage[usenames,dvipsnames]{color}
\usepackage[all]{xy}  
\usepackage{url}
\usepackage{amsfonts}
\usepackage{amssymb}
\usepackage{epsfig}
\usepackage{epstopdf}
\usepackage{bm}
\usepackage{balance}
\usepackage{graphicx}
\usepackage{subcaption}
\usepackage{footnote}
\usepackage{cancel}
\usepackage{algorithm,algorithmic}

\newtheorem{Lemma}{Lemma}

\newtheorem{Corollary}[Lemma]{Corollary}

\newtheorem{proposition}{Proposition}

 \usepackage{xcolor}

\DeclareMathOperator{\sinc}{sinc}

\newcommand{\qb}{{\bf b}}

\newcommand{\qH}{{\bf H}}
\newcommand{\qI}{{\bf I}}

\newcommand{\qW}{{\bf W}}

\newcommand{\mH}{{\bf{H}}}

\newcommand{\Kh}{\mathcal{K}_h}
\newcommand{\Kl}{\mathcal{K}_l}

\newcommand{\Htd}{{\bf{H}}^{\rm{TD}}}
\newcommand{\Htf}{{\bf{H}}^{\rm{TF}}}
\newcommand{\Hdd}{{\bf{H}}^{\rm{DD}}}

\newcommand{\FZF}{\mathtt{FZF}}
\newcommand{\PZF}{\mathtt{PZF}}
\newcommand{\MRT}{\mathtt{MRT}}

\newcommand{\BDH}{ \bar{\bf D}^{\rm H}}
\newcommand{\BD}{ \bar{\bf D}}
\newcommand{\D}{ {\bf D}}

\title{Hybrid OTFS/OFDM Design in Massive MIMO}

\author{Ruoxi Chong~\IEEEmembership{Student Member,~IEEE,} Mohammadali Mohammadi,~\IEEEmembership{Senior Member,~IEEE,}
Hien Quoc Ngo,\\~\IEEEmembership{Senior Member,~IEEE,}  Simon L. Cotton,~\IEEEmembership{Senior Member,~IEEE,} and  Michail Matthaiou,~\IEEEmembership{Fellow,~IEEE}
\thanks{The authors are with the Centre for Wireless Innovation (CWI), Queen's University Belfast, BT3 9DT Belfast, U.K.,
(email:\{rchong02, m.mohammadi, hien.ngo, simon.cotton, m.matthaiou\}@qub.ac.uk). Parts of this paper have appeared at the 2023 IEEE GLOBECOM conference~\cite{Ruoxi2023combine}.
}}

\begin{document}
\bstctlcite{IEEEexample:BSTcontrol}
\maketitle
\begin{abstract}
We consider a downlink (DL) massive multiple-input multiple-output (MIMO) system, where different users have different mobility profiles. 
To support this system, we categorize the users into two disjoint groups according to their mobility profile and implement a hybrid orthogonal time frequency space (OTFS)/orthogonal frequency division multiplexing (OFDM) modulation scheme.
Building upon this framework, two precoding designs, namely full-pilot zero-forcing (FZF) precoding and partial zero-forcing (PZF) precoding are considered. To shed light on the system performance, the spectral efficiency (SE) with a minimum-mean-square-error (MMSE)-successive interference cancellation (SIC) detector is investigated. Closed-form expressions for the SE are obtained using some tight mathematical approximations. To improve fairness among different users, we consider max-min power control for both precoding schemes based on the closed-form SE expression. However, by noting the large performance gap for different groups of users with PZF precoding, the per-user SE will be compromised when pursuing overall fairness. Therefore, we propose a weighted max-min power control scheme. By introducing a weighting coefficient, the trade-off between the per-user performance and fairness can be enhanced. Our numerical results confirm the theoretical analysis and reveal that with mobility-based grouping, the proposed hybrid OTFS/OFDM modulation significantly outperforms the conventional OFDM modulation for high-mobility users. 


\let\thefootnote\relax\footnotetext{This work is a contribution by Project REASON, a UK Government funded project under the Future Open Networks Research Challenge
(FONRC) sponsored by the Department of Science Innovation and Technology (DSIT). The work of M. Mohamamdi and M. Matthaiou was supported by the European Research Council (ERC) under the European Union’s Horizon 2020 research and innovation
programme (grant agreement No. 101001331). The work of  H.~Q.~Ngo
 was supported by the U.K. Research and Innovation Future
Leaders Fellowships under Grant MR/X010635/1, and a research grant from the Department for the Economy Northern Ireland under the US-Ireland R\&D Partnership Programme.}
\end{abstract}

\section{Introduction}
Beyond fifth-generation (B5G) wireless communication systems are envisioned to provide reliable communication services under various heterogeneous channel conditions~\cite{Matthaiou:COMMag:2021}. The currently deployed orthogonal frequency division multiplexing (OFDM) modulation has demonstrated great performance over the years due to its great resilience against time dispersion, achieved through the introduction of cyclic prefix (CP). However, as high-mobility scenarios have become an indispensable part of human life, with velocities reaching up to $500$ km/h on high-speed railways and around $900$ km/h on airplanes, wireless channels exhibit doubly dispersive manifestations in the time-frequency (TF) domain. More specifically, time dispersion is caused by the effects of multipath propagation, while frequency dispersion is caused by Doppler shifts. In such cases, the currently deployed OFDM modulation may break down because the significant Doppler spread induced by the high mobility can severely undermine the orthogonality between subcarriers. 

Different from OFDM modulation, orthogonal time frequency space (OTFS) modulation multiplexes the information symbols in the delay-Doppler (DD) domain. With the aid of the DD domain signal processing, the channel responses are relatively sparse and static~\cite{Hadani2017orthogonal,Wei2021magzine,Shuangyang2023Globecom,li2024fundamentals}. Furthermore, the symbol placement in the DD domain enables direct interaction between the information symbols and channel responses, resulting in a much simpler input-output relationship compared to that of the OFDM modulation in high-mobility channels~\cite{Li2021performance}. 
By invoking the two-dimensional (2D) inverse symplectic finite Fourier transform (ISFFT), each DD domain symbol spreads onto the whole TF domain, thus principally experiencing the entire perturbation of the TF channel over an OTFS frame. Therefore, OTFS modulation offers the potential of harnessing the full channel diversity~\cite{Li2021performance}. With all the mentioned advantages introduced by the OTFS modulation, many works have been done in this field from different aspects.

For example, the application of different multiple access (MA) schemes for OTFS systems has become a popular topic. Specifically, in~\cite{Ruoxi2022outage,Ruoxi2022achievable,Ruoxi2022SE} two orthogonal MA schemes were proposed, namely delay division multiple access and Doppler division multiple access, and the achievable rates for both schemes were discussed. The coexistence of non-orthogonal multiple access (NOMA) and OTFS was investigated in~\cite{zhiguoOTFSNOMA}, in which users with different mobility profiles were grouped together for the implementation of NOMA in both uplink (UL) and downlink (DL) transmission. Analytical results demonstrated that OTFS-NOMA improves the spectral efficiency (SE) and reduces latency~\cite{zhiguoOTFSNOMA,WenNOMA}. 

The potential of multiple-input multiple-output (MIMO) and massive MIMO technology to enhance the SE of OTFS systems has also been investigated. Specifically, Liu \emph{et al.}~\cite{Wang:JSAC:2020} proposed a path division MA scheme for both UL and DL transmission for a massive MIMO-OTFS architecture. Li \emph{et al.}~\cite{Muye:JSAC:2021} and Shi \emph{et al.}~\cite{Shi:TWC:2021} studied OTFS modulation for massive MIMO systems, with a focus on channel estimation. Shen \emph{et al.}~\cite{ShenChannel} proposed a 3D-structured orthogonal matching pursuit algorithm-based channel estimation technique for OTFS massive MIMO systems. The authors in~\cite{MM2021meet,MMmeetDownlink},  showed the tradeoff between the system performance and the signaling overhead for a cell-free massive MIMO system with OTFS modulation. A simple implementation of the DD domain Tomlinson-Harashima precoding for DL multiuser MIMO OTFS transmissions was proposed in~\cite{Li2023THP}. Saeid \emph{et al.}~\cite{Saeid2023ISAC} proposed a beam-space MIMO radar design to enable a joint communication and sensing system with OTFS modulation operating in millimeter-wave frequency bands.
There is also some work focused on the millimeter wave (mmWave) bands with MIMO-OTFS modulation. 
For example, the authors in~\cite{mmWave_2} proposed a two-stage framework to maximize the directional beamforming gains, while the authors in~\cite{Mu2022mmWave} proposed a joint channel estimation and data detection method using a message-passing algorithm.

Extensive literature indicates that OTFS modulation can provide more robust performance than OFDM in high-mobility channels~\cite{GiuseppeOTFS}.
Nevertheless, under the current OFDM system setup, introducing DD domain signal processing and applying OTFS modulation will entail some extra domain transformation processes, including an ISFFT and symplectic finite Fourier transform (SFFT), resulting in a higher computational complexity~\cite{Raviteja:TWC:2018}. In this context, combining OTFS and OFDM, by viewing OTFS as complementary to OFDM in high-mobility conditions, results in an interesting performance-complexity trade-off.

Despite the extensive literature on massive MIMO-OTFS systems, the combination of OTFS and OFDM, along with various precoding designs, has not been thoroughly studied in the massive MIMO space. 
To bridge this gap, in this paper, we consider a DL massive MIMO system with users having different mobility profiles and propose a hybrid OTFS/OFDM transmission protocol with different precoding designs. Specifically, we divide the users into two disjoint groups based on their mobility profile, namely high-mobility users (HM-UEs) and low-mobility users (LM-UEs). We apply OTFS modulation for HM-UEs and OFDM modulation for LM-UEs, while the precoding design is determined according to the system performance requirements and tolerable complexity. Two different precoding designs are considered at the base station (BS), referred to as full-pilot zero-forcing (FZF) and partial zero-forcing (PZF). The former scheme applies zero-forcing (ZF) for all users, completely suppressing inter-user interference at the cost of high computational complexity. On the other hand, the latter PZF scheme, which employs ZF for HM-UEs and maximum-ratio transmission (MRT) for LM-UEs, enables us to further balance between complexity and performance at the expense of inter-user interference for some users. With these two precoding schemes, we also address fairness among different users by implementing different power allocation designs at the BS. The main  contributions of this paper can be summarized as follows:
\begin{itemize}
    \item We discuss the frame design for OTFS modulation and compare it with that of the OFDM modulation. We derive an OTFS-equivalent matrix-form input-output relationship for the MIMO-OFDM system, with the consideration of adding CP and removing CP. 
    \item By looking into the considered systems' computational complexity, we propose and analyze FZF and PZF precoding for the massive MIMO system with hybrid OTFS/OFDM modulation.    
    We derive the complexity of the considered precoding schemes using the big $O$ function. We find that the complexity of PZF is dependent on the number of high-mobility users.
    To have a better understanding of the trade-off between complexity and performance, we give the SE of HM- and LM-UEs with different numbers of high-mobility users.
    \item Relying on the statistical channel state information (CSI) at the receiver side, we apply a minimum-mean-square-error successive interference cancellation (MMSE-SIC) detection and derive new analytical expressions for the DL per-user SE of HM- and LM-UEs for different precoding designs. Corresponding closed-form SE expressions are approximated. The tightness of our approximation is then verified by numerical results.
    \item With a more practical large-scale fading model, which incorporates correlated shadowing, we consider power allocation design at the BS to provide fairness among users. Max-min power allocation is first considered. Due to the substantial gap in the SE performance between HM-UEs and LM-UEs under PZF precoding design, max-min power allocation results in a significant performance loss for HM-UEs. Therefore, we propose a weighted max-min power allocation method to achieve a better trade-off between the per-user SE performance and fairness. In the case of PZF with a priority given to the LM-UEs, further user scheduling (USC) is considered. The simulation shows that around $20\%$ performance improvement in the $95\%$-likely SE can be achieved for LM-UEs with the help of the USC.
\end{itemize}

The rest of this paper is organized as follows: In Section~\ref{sec:sys model}, we first provide a brief overview of OTFS and compare it with OFDM. Then, we describe the system model for the proposed OTFS/OFDM system with different precoding schemes. In Section~\ref{sec:Perf}, we analyze the per-user SE with an MMSE-SIC detection and provide the closed-form SE expressions for different cases. Power allocation schemes are discussed in Section~\ref{sec:power}. Finally, the numerical results and some discussions are provided in Section V, followed by some concluding remarks in Section~\ref{sec:conc}.

\emph{Notations:} We use bold upper-case letters to denote matrices, and bold lower-case letters to denote vectors. The superscripts $(\cdot)^{\rm{H}}$ and $(\cdot)^{\rm{T}}$ denote the Hermitian transpose and transpose of a matrix, respectively; ${{{\bf{F}}_N}}$ denotes the normalized discrete Fourier transform (DFT) matrix of size $N\times N$; $\qI_M$ and $\boldsymbol{0}_{M\times N}$ represent the $M\times M$ identity matrix and zero matrix of size $M\times N$, respectively; 
``$ \otimes $" denotes the Kronecker product operator; 
$\textrm{det}(\cdot)$ and $\rm{Tr}(\cdot)$ denote the determinant and trace operations of a matrix, respectively; ${\rm{vec}}\left( \cdot \right)$ denotes the vectorization of a matrix;
$\| \cdot \|$ returns the norm of a matrix;
$\mathbb{E}\{\cdot\}$ denotes the statistical expectation. Finally,
$\Re$ and $\Im$ denote the real part and the imaginary part of a complex component, respectively. 
\section{System Model}~\label{sec:sys model}
In this section, we provide a concise system model for the considered OTFS/OFDM modulation with massive MIMO. 
\vspace{-0.7em}
\subsection{Preliminaries on OTFS Transmitters}
\begin{figure}
    \centering
    \includegraphics[width=0.9\linewidth]{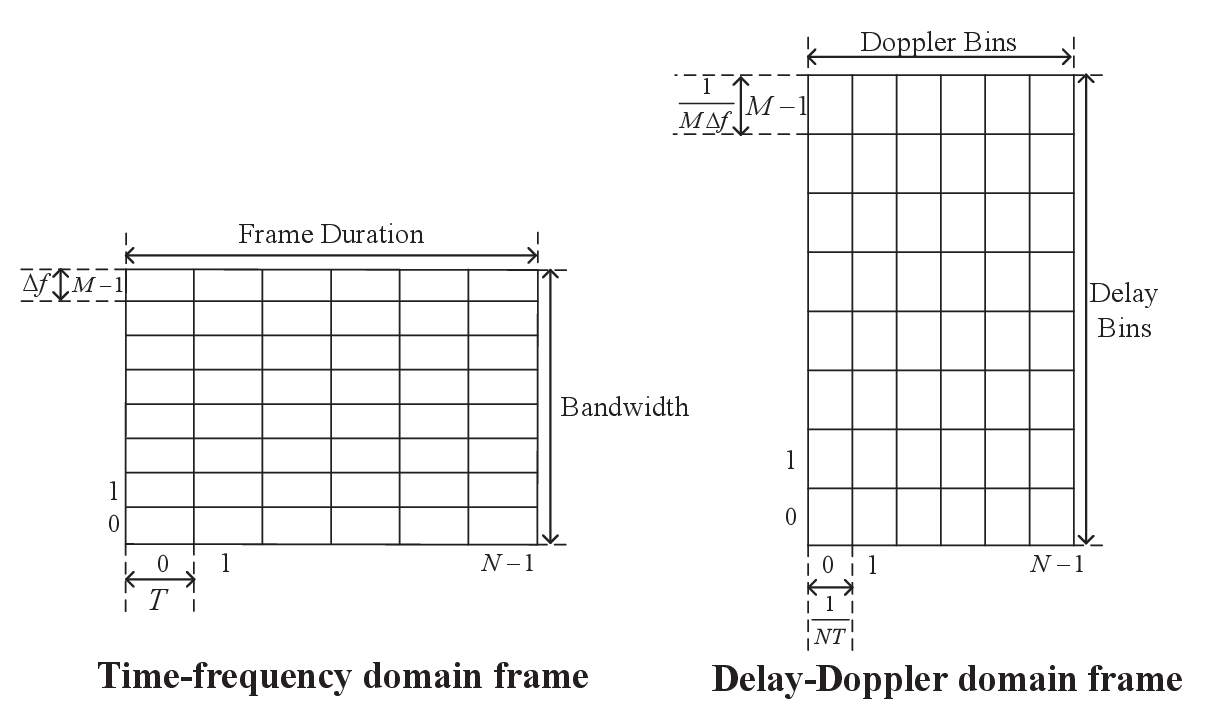}
    \caption{Frame comparison between OFDM and OTFS.}
    \label{fig:frame}
\end{figure}

We first consider a TF domain OFDM frame that occupies $M$ sub-carriers and $N$ time slots after adding the CP.  By applying the SFFT, an equivalent DD domain frame of size $M\times N$ for OTFS transmission can be obtained. Specifically, $M$ denotes the number of delay bins and $N$ is the number of frequency bins in the OTFS frame. The detailed frame design for OFDM and OTFS modulation is illustrated in Fig.~\ref{fig:frame}. Let us denote the sub-carrier frequency spacing as $\Delta f$, thus we have ${T}=1/\Delta f$ denoting the symbol duration. For a TF domain OFDM frame with a total bandwidth of $B_f=M\Delta f$ and a frame duration equal to $T_{f}=NT$ TF domain frame, the equivalent DD domain frame for OTFS can be viewed from Fig.~\ref{fig:frame}. We can see that, the \emph{delay resolution} and the \emph{Doppler resolution} for OTFS modulation are respectively determined by $1/(M\Delta f)$ and $1/(N{T})$, which means that with larger bandwidth and frame duration,  a more precise acquisition of the channel delay and Doppler can be obtained with OTFS modulation.

To gain a better understanding of the domain transformation in OTFS modulation, we show the process of obtaining the time-domain transmit signal with OTFS modulation. With the OTFS modulation, $MN$ number of users' information symbols ${\bf s}\in {\mathbb{A}}^{M N}$ will initially be mapped onto a two-dimensional 2D DD domain grid of size $M\times N$ for each frame, denoted as ${\bf{s}} \buildrel \Delta \over = \textrm{vec}\left( {\bf{S}} \right)$. Note that ${\mathbb{A}}$ represents an energy-normalized constellation set. Let us define the $(l,k)$-th element of ${\bf{S}}$, ${S\left[ {l,k} \right]}$, as the modulated pulse at the $k$-th Doppler and $l$-th delay grid point, for $0 \le k \le N-1, 0 \le l \le M-1$~\cite{Hadani2017orthogonal}. Then, the equivalent TF domain signal $X^{\rm TF}\!\left[ {n,m} \right]$ can be obtained by applying the ISFFT~\cite{Hadani2017orthogonal},
\vspace{-0.1em}
\begin{equation}
X^{\rm TF}\!\left[ {n,m} \right] = \frac{1}{{\sqrt {NM}} }\sum\limits_{k = 0}^{{N} - 1} {\sum\limits_{l = 0}^{{M} - 1} {S\left[ {k,l} \right]} } {e^{j2\pi \left( {\frac{{nk}}{{N}} - \frac{{ml}}{{M}}} \right)}}  .
\vspace{0.3em}
\end{equation}
With the TF domain transmitted symbols, the time domain transmit signal can then be obtained by using the conventional OFDM modulator, which can be achieved by an inverse fast Fourier transform (IFFT) module with the transmitter shaping pulse $g(t)$. The equivalent time domain transmit signal with OTFS modulation can then be denoted by
\vspace{-0.2em}
\begin{equation}
x^{\rm TD}\!\!\left( t \right) \!=\!\! \sum\limits_{n = 0}^{N - 1} {\sum\limits_{m = 0}^{M - 1}\!\! {X^{\rm TF}\!\left[ {m,n} \right]{g\left( {t - nT} \right)}{e^{j2\pi m\Delta f\left( {t - nT} \right)}}} }.\label{C4_OTFS_signal}
\vspace{0.3em}
\end{equation}
Notice that with OFDM modulation, only the second domain transformation in~\eqref{C4_OTFS_signal} is needed to obtain the time domain equivalent signal. 

\begin{figure}
\vspace{1.5em}
    \centering
    \includegraphics[width=1\linewidth]{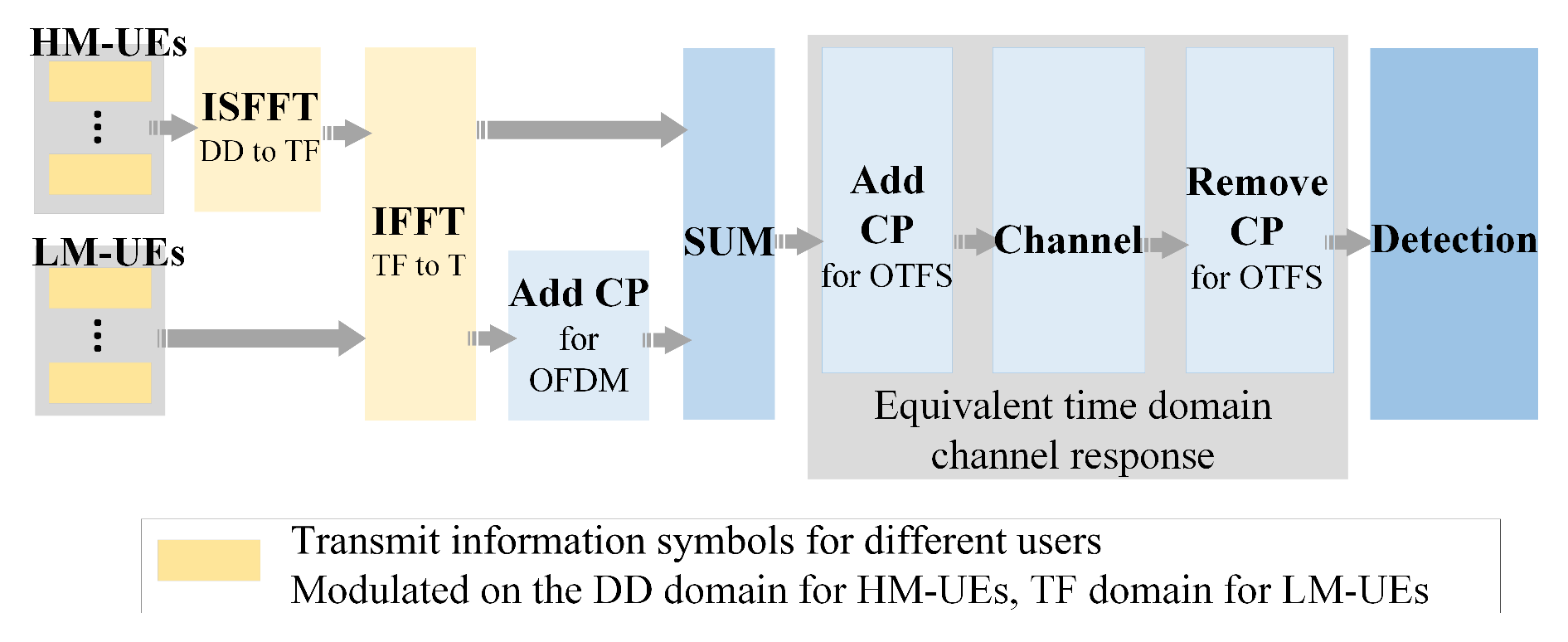}
    \caption{ 
        Illustration of the hybrid OTFS/OFDM modulation design.}
    \label{fig:system}
\end{figure}

\subsection{Input-Output Relationship}
In this paper, we consider a DL massive MIMO system consisting of one BS with $N_t$ antennas and $K$ single-antenna users. Furthermore, we assume a general scenario, in which users have heterogeneous mobility profiles e.g., some users are moving at high speeds, denoted by $\Kh \subset \{ 1,\ldots, K \}$, while others have low-mobility, denoted by $\Kl \subset \{ 1,\ldots, K \}$.{\footnote{Without loss of generality, the specific velocity threshold is not discussed here. The grouping is based on the relative high or low velocity.}} Note that $\Kh \cap \Kl = \varnothing$, $K_h=\left| \Kh \right|$, $K_l=\left| \Kl \right|$ and $K_h+ K_l = K$. 
To achieve a fair balance between complexity and performance, we apply classical OFDM modulation for LM-UEs, whilst OTFS is utilized for HM-UEs. 
The information symbols modulated onto the DD and TF domains will first be transferred into the time domain and then added together for transmission.
An illustration of the considered transmitter design is shown in Fig.~\ref{fig:system}. From Fig.~\ref{fig:system}, we can see that one additional ISFFT module and a different CP insertion mechanism are required at the transmitter side to implement OTFS modulation on the current OFDM system setup.

For HM-UEs, without loss of generality,  we consider a reduced-CP model in vector-form for OTFS modulation~\cite{Raviteja2019practical}. 
 Therefore, with the DD domain transmitted information signal for the ${k_h}$-th high-mobility user ${\bf{S}}_{k_h}\in {\mathbb{A}}^{M \times N}$, the TF domain equivalent signal can be denoted by~\cite{Wei2021transmitter}
\begin{equation}
{\bf{X}}^{\rm{TF}}_{k_h} = {{\bf{F}}_M}{\bf{S}}_{k_h} {\bf{F}}_N^{\rm{H}}.
\label{TF_symbol_matrix}
\vspace{0.3em}
\end{equation}
Then, by considering a rectangular pulse is used for the transmitter shaping pulse, the time domain transmitted symbol matrix can be obtained by~\cite{Wei2021transmitter}
\begin{equation}
{\bf{X}}^{{\rm{TD}}}_{k_h} = {{\bf{I}}_M}{\bf{F}}_M^{\rm{H}}{\bf{X}}^{\rm{TF}}_{k_h} = {\bf{S}}_{k_h} {\bf{F}}_N^{\rm{H}}.
\label{TD_symbol_matrix}
\vspace{0.3em}
\end{equation}
According to~\eqref{TD_symbol_matrix}, the equivalent time domain symbol vector for the $k_h$-th user,  ${\bf{x}}_{k_h}^{\rm TD}$, can be obtained as~\cite{Wei2021transmitter}
\begin{align}
  {\bf{x}}_{k_h}^{\rm TD} 
  \buildrel \Delta \over = {\rm{vec}}\left( {\bf{X}}^{{\rm{TD}}}_{k_h} \right)
  &= \left( {{{\bf{F}}_N^{\rm H}} \otimes {{\bf{I}}_M}} \right){\bf{s}}_{k_h} .
\label{x_kh_TD}
\vspace{0.3em}
\end{align}

To have a consistent system, we allocate each LM-UE with the same bandwidth and frame duration as the HM-UEs. Hence, to have an OFDM transmission occupying the $M \times N$ TF domain resource block, for each LM-UE, $L_d\buildrel \Delta \over=M-L_{\rm CP}$ information symbols will be sent for one symbol duration, with a $N$ total symbol duration for one frame. Therefore, by considering the TF domain information symbol vector ${\bf{s}}_{k_l}$ of length $L_dN$, the equivalent time domain sequence can be obtained by applying an IFFT,  given by 
 \begin{align}	
 {\Bar{\bf{x}}}_{k_l}^{\rm TD}
 &= \left( {{\bf{I}}_{N}\!\otimes\! {\bf{F}}^{\rm H}_{L_d}} \right) {\bf{s}}_{k_l}.
 \label{x_kl_TD}
 \vspace{0.3em}
 \end{align}
After applying the CP, the time domain transmit symbol can be represented as 
  \begin{align}	
 {\bf{x}}_{k_l}^{\rm TD}
 &=\left (  {\bf I}_{N}\!\otimes\!{{\bf A}_{\rm CP}}  \right ) {\Bar{\bf{x}}}_{k_l}^{\rm TD}
 =\left (  {\bf I}_{N}\!\otimes\!{{\bf A}_{\rm CP}{\bf{F}}^{\rm H}_{L_d}}  \right )  {\bf{s}}_{k_l},
 \vspace{0.3em}
 \end{align}
where ${\bf A}_{\rm CP}=[{\bf G}_{\rm CP}~ {\bf I}_{L_{d}}]^{\rm T}$ is the CP addition matrix of size $M\times L_d$, and ${\bf G}_{\rm CP}$ contains the last $L_{\rm CP}$ columns of ${\bf I}_{L_d}$.

Based on~\eqref{x_kh_TD} and~\eqref{x_kl_TD}, we further apply precoding and power allocation in the time domain for each user. The time domain transmitted signal sent by the BS for each frame can then be denoted by
\begin{align}
		{\bf{x}}^{\rm TD} 
		\!\!\!=\!  \sum\nolimits_{k_h\in \Kh}  \!\!\!\sqrt{\eta _ {k_h}} {\bf W}_{k_h} {\bf{x}}_{k_h}^{\rm TD} \!+\!  \sum\nolimits_{k_l\in \Kl}  
  \!\!\!\sqrt{\eta _ {k_l}} {\bf W}_{k_l} {\bf{x}}_{k_l}^{\rm TD},
  \vspace{0.3em}
\end{align}
where ${\eta _ {k_h}}$ and ${\eta _ {k_l}}$ are the power allocation coefficients for the $k_h$-th and $k_l$-th user; ${\bf W}_{k_h}$ and ${\bf W}_{k_l}$ are the precoding matrices of size $N_tMN\times MN$ for the $k_h$-th and $k_l$-th user. The precoding will be done on the RF chain, and its detailed structure will be exploited later.

We assume that the channel has perfect reciprocity and a total of $P$ independent resolvable paths exist between the BS and each user. Furthermore, we assume that the BS antenna is a uniform linear array with half wavelength inter-element spacing, and define $\phi_{k(i)}$ as the angle of arrival for the $i$-th resolvable path. The steering vector ${\bm \theta} _{k(i)}$ for the $i$-th path of size $1\times N_t$ is denoted\footnote{ Note that this model and the proposed communication protocols can be easily adapted to a three-dimensional model, by considering a steering vector with both zenith and azimuth angles.}
 by ${\bm \theta} _{k(i)}= \big[ 1, {\rm exp}(-j\pi \left(1\right )\sin \phi_{k(i)} ),\ldots , {\rm exp}(-j\pi \left( N_t -1\right )\sin \phi_{k(i)} )\big]$.
By considering the reduced-CP structure for OTFS transmission~\cite{Raviteja2019practical}, a CP block of length $L_{\rm CP}$ is inserted at the beginning of the whole frame in the time domain. Therefore, a total $(MN+L_{\rm CP})$-length data is transmitted for one frame, and the CP will be removed at the receiver.
The equivalent time domain channel response between the BS and the $k$-th user can be modeled as~\cite{Muye:JSAC:2021}
\begin{align}
\mH_{k}^{\rm TD}
 &=\!\! \sqrt{\beta _{k}}\!\sum\nolimits_{i=1}^{P} \!{\bm \theta} _{k(i)}\! \otimes \! \left({h_{k (i)}} \! {\bm{\Pi }}^{l_{k (i)}} \!{\bm{\Delta}}\!^{k_{k (i)}}\right),
\label{MIMOH}
\vspace{0.3em}
\end{align}
where $h_{k(i)}$ is the small-scale fading coefficient of the $i$-th path, which follows the Gaussian distribution with zero mean and $1/(2P)$ variance per real dimension; ${\bf{\Pi }}$ is a permutation matrix (forward cyclic shift) of size $MN\times MN$ characterizing the delay effect, i.e., $\boldsymbol{\Pi} =\mathrm{circ}\{[0,1,0,\ldots,0]^{\rm T}_{MN\times 1}\}$, and ${\bm{\Delta}}=\textrm{diag}\{{\alpha}^0,{\alpha}^1,\ldots,{\alpha}^{MN-1}\} $ is a diagonal matrix characterizing the Doppler effect with ${\alpha} \buildrel \Delta \over = {e^{\frac{{j2\pi }}{{MN}}}}$~\cite{Raviteja2019practical}. Furthermore, the terms  $l_{k (i)}$ and $k_{k (i)}$ in~\eqref{MIMOH} are the indices of delay and Doppler associated to the $i$-th path, respectively;\footnote{Note that \eqref{MIMOH} gives a close approximation when the system has fractional Doppler indices~\cite{ViterboBook}.} $\beta _{k}$ is the large-scale fading coefficient for the $k$-th user. 
Without loss of generality, in this paper, we consider integer delay and fractional Doppler. Since the sampling time $1/M \Delta f$ is usually sufficiently small, the impact of fractional delay is neglected in this paper~\cite{Raviteja:TWC:2018}.

Therefore, the received signal in the time domain for the $k$-th user is denoted by
\begin{align}
		{\bf y}^{\rm TD}_{(k)} 
		&= \sum\nolimits_{k=1}^{K} {\sqrt{\rho \eta_k}} \Htd _{k} {\bf W}_{k}{\bf{x}}_{k}^{\rm TD}+ {{\bf z}_{k}},
		\label{inout}
\vspace{0.3em}
\end{align}
where ${{\bf z}_{k}}$ is the additive white Gaussian noise (AWGN) sample vector, with ${\mathbb E}\left\{ {{\bf z}_{k}} {{\bf z}_{k}}^{\rm H} \right\}={\bf I}_{ MN}$, while $\rho$ is the normalized signal-to-noise ratio (SNR).

For notation simplicity, we define the DD domain and TF domain equivalent channel matrices as follows
\vspace{0.5em}
\begin{align}
\mH_{k}^{\rm DD}
&= ({{{\bf{F}}_N} \otimes {{\bf{I}}_M}}) \Htd _{k} \left( {{\bf I}_{N_t} \otimes{{\bf{F}}_N^{\rm H}} \otimes {{\bf{I}}_M}} \right),\\\vspace{0.2em}
\mH_{k}^{\rm TF}
&=({ {{\bf{I}}_N} \otimes  {{\bf{F}}_M} }) \Htd _{k} \left( {{\bf I}_{N_t} \otimes{{\bf{I}}_N} \otimes {{\bf{F}}^{\rm H}_M}} \right).
\vspace{1em}
\end{align}
Hence, for the $k_h$-th HM-UE, the equivalent DD domain received signal for the $k_h$-th HM-UE is shown in~\eqref{inout_kh} at the top of the next page.
\begin{figure*}
{\small
\begin{align}
{\bf y}^{\rm DD}_{(k_h)} &\!= \!({{{\bf{F}}_N} \otimes {{\bf{I}}_M}}) {\bf y}^{\rm TD}_{(k_h)} 
\!=\!\underbrace{  {\sqrt{\rho \eta_{k_h}}}   \Hdd _{k_h}   {\bf W}_{k_h} {\bf{s }}_{k_h} }_{\text{Desired signal}}
\!+\!\!\underbrace{\!\! \sum_{k'_h\in \Kh,\atop k'_h\neq k_h} \!\!\!\!\! {\sqrt{\rho \eta_{k'_h}}}   \Hdd _{k_h} {\bf W}_{k'_h}  {\bf{s }}_{k'_h}}_{\text{Intra-group interference}} \! 
+ \underbrace{ \!\!\!\!\sum_{k'_l\in \Kl} \!\!\!\! {\sqrt{\rho \eta_{k'_l}}}  ({{{\bf{F}}_N} \!\otimes\! {{\bf{I}}_M}}) \Htd _{k_h} \left( {{\bf I}_{N_t} \!\otimes\!{{\bf{I}}_N} \!\otimes\! {{\bf{F}}^{\rm H}_M}} \right)  {\bf W}_{k'_l}  {\bf{s }}_{k'_l} }_{\text{Inter-group interference}} +{{\bf z}_{k_h}},
		\label{inout_kh}
\end{align}
}
{\small
\begin{align}
{\bf y}^{\rm TF}_{(k_l)} 
&\!=\! ({ {{\bf{I}}_N} \!\otimes\!  {{\bf{F}}_M} }) {\bf y}^{\rm TD}_{(k_l)} 
\!=\!\underbrace{{\sqrt{\rho \eta_{k_l}}}   \Htf _{k_l} {\bf W}_{k_l} {\bf{s }}_{k_l} }_{\text{Desired signal}}
\!+\!\underbrace{\!\!\!\!\sum_{k'_l\in \Kl,\atop k'_l\neq k_l}\!\! {\sqrt{\rho \eta_{k'_l}}}   \Htf _{k_l}   {\bf W}_{k'_l} {\bf{s }}_{k_l'} }_{\text{Intra-group interference}} 
+\!\! \underbrace{\sum_{k'_h\in \Kh}\!\! {\sqrt{\rho \eta_{k'_h}}} ({ {{\bf{I}}_N} \otimes  {{\bf{F}}_M} }) \Htd _{k_l} \left( {{\bf I}_{N_t} \otimes{{{\bf{F}}_N^{\rm H}} \!\otimes\! {{\bf{I}}_M}}} \right)   {\bf W}_{k'_h}  {\bf{s }}_{k'_h} }_{\text{Inter-group interference}} \!+\!{{\bf z}_{k_l}}.
\label{inout_kl}
\end{align}
}
\centering
\rule{\textwidth}{0.3mm}
\end{figure*}

For the $k_l$-th LM-UE, the equivalent TF domain received signal can be obtained by first removing the CP using ${\bf R}_{\rm CP}$ in the time domain, and then applying a fast Fourier transform (FFT) for the domain transformation. Therefore, by substituting~\eqref{x_kh_TD} and~\eqref{x_kl_TD}, the input-output relationship for the equivalent TF domain received signal is represented in~\eqref{inout_kl} at the top of the next page. Notice that the CP removal matrix ${\bf R}_{\rm CP}$ in~\eqref{inout_kl} is of size $L_d\times M$, and it equals to ${\bf I}_M$ after removing the first $L_{\rm CP}$ rows.

\vspace{-1em}
\section{Performance Analysis}~\label{sec:Perf}
We assume that the considered transmission is inside a stationarity region, where the effective channel is wide-sense stationary uncorrelated scattering (WSSUS) and deterministic in the DD domain~\cite{WSSUS}.
Therefore, we consider an MMSE-SIC detector with perfect CSI known at the transmitter side~\cite{Li2023THP} and analyze the SE performance of different precoding designs.
Note that with the consideration of the perfect CSI, our analysis provides an achievable upper bound of the system performance.
According to~\cite{SE}, with an input-output relationship as ${\bf y}_k=\sum_{k'=1}^{K} {\bf H}_{k} {\bf W}_{k'} {\bf s}_{k'} $, the DL achievable SE can be obtained as
\begin{align}~\label{eq:SEk}
{\rm{SE}}_k
&=\alpha_{\rm{SE}} {\log _2}\det \left( {\bf{I}}_{MN} +  \BDH_{k k} \left( {\bf {\bf \Psi}}_k \right)^{-1} \BD_{k k} \right),
\vspace{0.3em}
\end{align}
where $\BD_{k k}=\mathbb{E}\left\{ {\D}_{k k} \right\}$, and 
\begin{subequations}
	\begin{align}
		\D_{k k} &=  \qH_{k}   {\bf W}_{k} ,\\
		\D_{k k'} &=  {\qH}_{k}  {\bf W}_{k'}  , \\
		{\bf \Psi}_k &= {\bf I}_{MN} + \mathbb E \Big\{\sum\nolimits^{K}_{k'=1} \D_{k k'} \D_{k k'}^{\rm H}  \Big\} - \BD_{k k} \BDH_{k k} \label{21c},
	\end{align}  
\end{subequations}
where $\alpha_{\rm{SE}}$ is a normalization coefficient, and in our case we have $\alpha_{\rm{SE}}=\frac{1}{MN+L_{\rm CP}}$. This DL achievable SE will be applied for our later discussion, and we notice that~\eqref{eq:SEk} results in a tight approximation to the real system due to the channel hardening effect provided by massive MIMO~\cite{HienBook}.
Note that we assume the BS has sufficient computing and memory resources for the precoding and power allocation.

\vspace{-1em}
\subsection{FZF precoding}
Let us first look into the FZF precoding scheme for all users. 
With the grouping method based on the users' mobility profile, we further define ${\mH}^{\rm FZF}$
\noindent {\small$=\![(\Htd _{1})\!^{\rm T}\!,(\Htd _{2})\!^{\rm T}\!,\ldots, (\Htd_{K_h})\!^{\rm T}\!,(\Htd _{1})\!^{\rm T}\!,(\Htd _{2})\!^{\rm T}\!,\ldots,(\Htd_{K_l})\!^{\rm T}]^{\rm T}$}. 
Note that the size of ${\mH}^{\rm FZF}_{\rm H}$  is $KMN\times N_tMN$. 
Then, for the $k_h$-th HM-UE, the precoding matrix is designed as
\begin{equation}
\qW_{k_h}^\FZF=\alpha^{\FZF}_{k_h} ({\mH}^{\rm FZF})^{\rm H} \left (  {\mH}^{\rm FZF} ({\mH}^{\rm FZF})^{\rm H} \right )^{-1} {\bf B}_{k_h} ,
  \label{FZF}
\end{equation}
with
\vspace{-0.5em}
\begin{equation}
\alpha^{\FZF}_{k_h} \!= \!\frac{{\sqrt{MN}}}{\sqrt{\mathbb E \left\{ \left\| ({\mH}^{\rm FZF})^{\rm H} \left (  {\mH}^{\rm FZF} ({\mH}^{\rm FZF})^{\rm H}\right )^{-1} {\bf B}_{k_h} \right\| ^2 \right\} }},
\end{equation}

\noindent where ${\bf B}_{k_h}=[\big ( {{\bf{b}}_{K_h}^{(k_h)}} \otimes {\bf I}_{MN} \big),  {\bf 0}_{MN \times K_l M N}]^{\rm T}$ is of size $KMN \times MN$, and ${{\bf{b}}_{K_h}^{(k_h)}}$ is a row vector of length $K_h$, with only the $k_h$-th entry being one and others being zero. Moreover, with $ {\bf B}_{k_h}^{\rm H} {\mH}^{\rm FZF} = \Htd _{k_h}$, we can see that ${\bf B}_{k_h}$ helps to pick out the $k_h$-th matrix from the block matrix ${\mH}^{\rm FZF}_{\rm H}$. 
Note that $\alpha^{\FZF}_{k_h}$ is the normalization coefficient, with
\begin{align}
&{{\mathbb E \Big\{ \big\| ({\mH}^{\rm FZF})^{\rm H} \left (  {\mH}^{\rm FZF} ({\mH}^{\rm FZF})^{\rm H}\right )^{-1} {\bf B}_{k_h} \big\| ^2 \Big\} }}\notag\\
&=\mathbb E \Big\{{\rm Tr} \big[{\bf B}_{k_h}^{\rm H} \left (  {\mH}^{\rm FZF} ({\mH}^{\rm FZF})^{\rm H} \right )^{-1} {\mH}^{\rm FZF} ({\mH}^{\rm FZF})^{\rm H} \notag \\
&~~~~\times\left ( {\mH}^{\rm FZF}({\mH}^{\rm FZF})^{\rm H} \right )^{-1} {\bf B}_{k_h}  \big] \Big\} \notag\\
&=\frac{1}{K}{{\mathbb E \Big\{  {\rm Tr} \big[  \left (  {\mH}^{\rm FZF}({\mH}^{\rm FZF})^{\rm H}  \right )^{-1}    \big ] \Big\} }}.
\label{alpha_FZF}
\end{align}

\noindent Therefore, the normalization coefficient $\alpha^{\FZF}_{k_h}$ can be further expressed as{\footnote{The normalization coefficients are assumed to be known at the BS, as they can be considered as constant within a coherence block.}}
\vspace{-0.5em}
\begin{align}
\alpha^{\FZF}_{k_h}
=\frac{\sqrt{KMN}}{\sqrt{{\mathbb E \Big\{ {\rm Tr} \big[  \left (  {\mH}^{\rm FZF}({\mH}^{\rm FZF})^{\rm H}  \right )^{-1}    \big ] \Big\} }}}.
\end{align}
 
Similarly, for the $k_l$-th LM-UE, the precoding design is represented by  
\vspace{-0.3em}
\begin{equation}
\qW_{k_l}^\FZF=\alpha^{\FZF}_{k_l} 
		({\mH}^{\rm FZF})^{\rm H} \left (  {\mH}^{\rm FZF} ({\mH}^{\rm FZF})^{\rm H} \right )^{-1} {\bf B}_{k_l},
  \label{FZF_L}
   \vspace{0.3em}
\end{equation}
with $\alpha_{\FZF}\triangleq\alpha^{\FZF}_{k_l} =\alpha^{\FZF}_{k_h}$ and ${\bf B}_{k_l}=[{\bf 0}_{MN \times K_h M N}, \big ( {{\bf{b}}_{K_l}^{(k_l)}} \otimes {\bf I}_{M N} \big)]^{\rm T}$ is of size $KMN\times MN$. 
Due to the structure of ${\bf B}_{k_h}$ and ${\bf B}_{k_l}$, we can easily prove that ${\bf B}_{k}^{\rm H} {\bf B}_{k}={\bf I}_{MN}$, ${\bf B}_{k}^{\rm H} {\bf B}_{k'}={\bf 0}_{MN}$ with $k\neq k'$.

\begin{proposition}~\label{theorem:FZF:LM}
With FZF, the SE for the $k_h$-th HM and the $k_l$-th LM-UE can be obtained in closed-form as
\begin{subequations}
	\begin{align}
		{\rm{SE}}_{k_h}^\FZF&= \alpha_{\rm SE} MN {\log _2} \left( 1 +  \alpha_{\FZF}^{2} \rho\eta_{k_h}  \right), \label{SE_HFZF} \\
		{\rm{SE}}_{k_l}^\FZF&=\alpha_{\rm SE} L_d N {\log _2} \left( 1 +  \alpha_{\FZF}^{2} \rho\eta_{k_l} \right).
     \label{SE_LFZF}
	\end{align}
 \end{subequations}
\end{proposition}
\begin{proof}
    See Appendix~\ref{APX:theorem:FZF:LM}.
\end{proof}
From Proposition 1, we observe that by using FZF, all the intra-group and inter-group interference can be canceled at the cost of high computational complexity for both HM and LM-UEs. The main performance difference comes from the different levels of overhead. In this context, for a frame of length $(MN+L_{\rm CP})$, an $L_{\rm CP}$-length CP is added for the HM-UEs with OTFS, while an $L_{\rm CP}\times (N+1)$-length CP is considered for the LM-UEs with OFDM. Therefore, compared to the OTFS counterpart, a larger CP overhead is required for OFDM modulation, which results in a lower SE for LM-UEs. 
Moreover, note that higher reliability can be provided by OTFS modulation, due to its potential to achieve full diversity~\cite{Li2021performance}.

\vspace{-1em}
\subsection{PZF Precoding}
Implementing FZF requires high complexity, and, thus, we now consider a precoding design with lower complexity. We consider PZF precoding for HM-UEs to suppress inter-group interference. Subsequently, maximum ratio transmission (MRT) precoding is applied to the LM-UEs due to its low complexity and good performance, especially in low SNR regimes.
For the HM-UEs, the PZF precoding matrix $\qW_{k_h}^\PZF$ can be expressed as
\begin{equation}
		\qW_{k_h}^\PZF\!=\!\alpha^{\PZF}_{k_h} 
		(\mH^\PZF)^{\rm H} \left (  \mH^\PZF (\mH^\PZF)^{\rm H} \right )^{\!-1} \big ( {\qb_{K_h}^{(k_h)}} \!\otimes\! \qI_{MN} \big) ,
  \label{PZF}
\end{equation}
where $\mH^\PZF=[(\Htd _{1})^{\rm T},(\Htd _{2})^{\rm T},\ldots,(\Htd_{K_h})^{\rm T}]^{\rm T}$ with a size of $K_hMN\times N_tMN$, and
\begin{align}~\label{eq:alphaPZF}
\alpha^{\PZF}_{k_h}\!=\! \frac{{\sqrt{MN}}}{\sqrt{\mathbb E \Big\{ \big\| (\mH^\PZF)^{\rm H} \left ( \mH^\PZF (\mH^\PZF)^{\rm H}\right )^{-1} \!\big ( {{\bf{b}}_{K_h}^{(k_h)}} \otimes {\bf I}_{MN} \big) \big\| ^2 \Big\} }},
\vspace{0.3em}
\end{align} 
is the normalization coefficient.  
As in the previous case, we have
\begin{align}
&\mathbb E \Big\{ \big\| (\mH^\PZF)^{\rm H} \left ( \mH^\PZF (\mH^\PZF)^{\rm H}\right )^{-1}  ( {{\bf{b}}_{K_h}^{(k_h)}} \otimes {\bf I}_{MN} ) \big\| ^2 \Big\}\notag\\
&=\mathbb E \Big\{{\rm Tr} \Big[\big ( ({{\bf{b}}_{K_h}^{(k_h)}})^{\rm H} \otimes {\bf I}_{MN} \big)  \left (  {\mH}^{\PZF} ({\mH}^{\PZF})^{\rm H} \right )^{-1} {\mH}^{\PZF}\notag \\
&~~~~\times({\mH}^{\PZF})^{\rm H} \left ( {\mH}^{\PZF} ({\mH}^{\PZF})^{\rm H} \right )^{-1}  ( {{\bf{b}}_{K_h}^{(k_h)}} \otimes {\bf I}_{MN} )  \Big] \Big\} \notag\\
&= \frac{1}{K_h}{{\mathbb E \big\{ {\rm Tr} \big[  \left (  {\mH}^{\PZF}({\mH}^{\PZF})^{\rm H}  \right )^{-1}    \big ] \big\} }}.
\vspace{0.5em}
\end{align}
Therefore, the normalization coefficient~\eqref{eq:alphaPZF}  can be further derived as
\vspace{-0.5em}
\begin{align}
\alpha^{\PZF}_{k_h} 
=\frac{\sqrt{MNK_h}}{\sqrt{{{\mathbb E \big\{ 
{\rm Tr} \big[  \left (  {\mH}^{\PZF}({\mH}^{\PZF})^{\rm H}  \right )^{-1}    \big ] \big\} }}}}.
\vspace{0.5em}
\end{align}
For LM-UEs, to apply the MRT precoding, we have 
\begin{equation}
		\qW_{k_l}^\MRT=\alpha^{\MRT}_{k_l}{(\Htd _{k_l})}^{\rm H},
  \label{W_MRT}
  \vspace{0.3em}
\end{equation}
where $\alpha^{\MRT}_{k_l}=\frac{\sqrt{MN}}{\sqrt{\mathbb E \big\{ \big\| \Htd _{k_l} \big\| ^2 \big\} }}$, with
\begin{align}
&\mathbb E \big\{ \big\| \Htd _{k_l} \big\| ^2 \big\} 
=\mathbb E\left\{{\rm Tr}\left [ \Htd _{k_l} (\Htd  _{k_l})^{\rm H} \right ]\right\}
\notag\\
&=\beta _{k_l} \mathbb{E}\bigg\{ {\rm Tr} \Big[ \sum_{i=1}^{P} \mathbb E \left\{ {\bm \theta} _{k_l(i)} {\bm \theta} _{k_l(i)}^{\rm H} \right\} \otimes \mathbb E \left\{ {\bf H}^{\rm TD} _{k_l(i)}  ({\bf H}^{\rm TD} _{k_l(i)})^{\rm H}  \right\} \notag\\
&~~+ \sum_{i=1}^{P} \sum_{j=1, j\neq i}^{P}\mathbb E \left\{ {\bm \theta} _{k_l(i)} {\bm \theta} _{k_l(j)}^{\rm H} \right\} \otimes \mathbb E \left\{ {\bf H}^{\rm TD} _{k_l(i)}  ({\bf H}^{\rm TD} _{k_l(j)})^{\rm H}  \right\}  \Big ] \bigg\} \notag\\
&\stackrel{(a)}=\beta _{k_l}  \mathbb{E}\bigg\{ {\rm Tr} \Big[  \sum_{i=1}^{P} \mathbb E \left\{ {\bm \theta} _{k_l(i)} {\bm \theta} _{k_l(i)}^{\rm H} \right\} \otimes \mathbb E \left\{ {h_{k_l (i)}}{h_{k_l (i)}^{\rm H}}  {\bf I}_{MN} \right\} \Big ] \bigg\} 
 \notag\\
&=\beta _{k_l} N_tMN,
\label{a_n_mrt}
\vspace{1em}
\end{align}

\noindent where $(a)$ in~\eqref{a_n_mrt} follows the fact that the zero-mean channel coefficients for different paths are independent of each other. Therefore, the normalization coefficient becomes,
\begin{align}
\alpha^{\MRT}_{k_l} 
=\frac{1}{\sqrt{\beta _{k_l} N_t}}.
\label{alpha_MRT}
\end{align}
\begin{proposition}~\label{theorem:PZF}
 With PZF precoding, the SE for the $k_h$-th HM-UE can be derived as
\begin{align}
{\rm{SE}}_{k_h}^\PZF&=\alpha_{\rm{SE}} {\log _2}\det \Big( {\bf I}_{MN}+\alpha_{\FZF}^2 {\rho\eta_{k_h}} \Big ( {\bf I}_{MN} \notag\\
&\hspace{2em}+ \sum_{k_l'\in\Kl} \mathbb E \Big\{ \D_{k_h k'_l} \D_{k_h k'_l}^{\rm H}  \Big\} \Big)^{-1} {\bf I}_{MN} \Big),
\label{SE_PZF_1}
\end{align}
with
\vspace{-0.1em}
\begin{align}
&\mathbb E \left\{ \D_{k_h k_l'} \D_{k_h k_l'}^{\rm H}  \right\} 
=(\alpha^{\MRT}_{k'_l}) ^2\rho\eta_{k_l'} ({\bf{F}}_N \otimes {\bf{I}}_M) \mathbb E \Big\{\Htd _{k_h}   (\Htd _{k_l'})^{\rm H}  \notag \\
&\hspace{2em}\times\left (  {\bf I}_{N}\otimes{\bf A}_{\rm CP} {\bf A}_{\rm CP}^{\rm H} \right )   \Htd _{k_l'} (\Htd _{k_h})^{\rm H}\Big\} ({\bf{F}}_N^{\rm H} \otimes {\bf{I}}_M).
\label {PZF_H_1}
\end{align}
\end{proposition}
\begin{proof}
See Appendix~\ref{Ptheorem:PZFMRT}.    
\end{proof}
To further simplify~\eqref{SE_PZF_1}, let us focus on the matrix ${{\bf A}_{\rm CP} {\bf A}_{\rm CP}^{\rm H}}$. 
Due to the special structure of ${\bf A}_{\rm CP}$, its diagonal entries are 1, while most off-diagonal entries are $0$, except for some that are $1$. The number of these entries is dependent on $L_{\rm CP}$. For example, with $M=4$ and $L_{\rm CP}=1$, we have
\begin{align}
    {\bf A}_{\rm CP} {\bf A}_{\rm CP}^{\rm H}
    \!\!=\!\!  \left[ {\begin{array}{*{20}{c}}
      1 & 0 & 0 & 1\\
      0 & 1 & 0 & 0\\
      0 & 0 & 1 & 0\\
      1 & 0 & 0 & 1
      \end{array}} \right].
\end{align}
As $L_{\rm CP}$ is usually around $20\%$ of $M$ in the common OFDM system, for simplicity, we use the approximation that ${\bf A}_{\rm CP} {\bf A}_{\rm CP}^{\rm H} \approx {\bf I}_M$.
Therefore,~\eqref{PZF_H_1} can then be approximated as
\begin{align}
\mathbb E \Big\{ \D_{k_h k_l'} \D_{k_h k_l'}^{\rm H}  \Big\} 
&\approx(\alpha^{\MRT}_{k'_l}) ^2\rho\eta_{k_l'} ({\bf{F}}_N \otimes {\bf{I}}_M) \mathbb E \Big\{\Htd _{k_h}   (\Htd _{k_l'})^{\rm H}   \notag \\
&\hspace{2em} \times\Htd _{k_l'} (\Htd _{k_h})^{\rm H}\Big\} ({\bf{F}}_N^{\rm H} \otimes {\bf{I}}_M),
\label {PZF_H_2}
\end{align}
with which, we have
{\small
\begin{align}
&\mathbb E \left\{\Htd _{k_h}   (\Htd _{k_l'})^{\rm H}   \Htd _{k_l'} (\Htd _{k_h})^{\rm H}\right\} \notag\\
&\stackrel{(a)}{=} \beta_{k_h}\beta_{k'_l} \sum_{i=1}^{P} \sum_{j=1}^{P} \sum_{m=1}^{P} \sum_{n=1}^{P} \!\mathbb E \left\{ {\bm \theta} _{k_h(i)} {\bm \theta} _{k_l'(j)}^{\rm H} {\bm \theta} _{k_l'(m)} {\bm \theta} ^{\rm H}_{k_h(n)}\right\}  \notag \\
&~~~\times\mathbb E \left\{ {\bf H}^{\rm TD} _{k_h(i)}  ({\bf H}^{\rm TD} _{k_l'(j)})^{\rm H} {\bf H}^{\rm TD} _{k_l'(m)}  ({\bf H}^{\rm TD} _{k_h(n)})^{\rm H} \right\}   \notag\\
&\stackrel{(b)}{=} \beta_{k_h}\beta_{k'_l} \sum_{i=1}^{P} \sum_{j=1}^{P} \!\mathbb E \left\{ {\bm \theta} _{k_h(i)} {\bm \theta} _{k_l'(j)}^{\rm H} {\bm \theta} _{k_l'(j)} {\bm \theta} ^{\rm H}_{k_h(i)}\right\}  \notag \\
&~~~\times \mathbb E \left\{ h_{k_h(i)} h_{k_l(j)}^* h_{k_l(j)} h_{k_h(i)}^*  \right\} {\bf I}_{MN}  \notag\\
&=\frac{\beta_{k_h}\beta_{k'_l}}{P^2}\sum_{i=1}^{P} \sum_{j=1}^{P} 
\mathbb E \left\{ {\bm \theta} _{k_h(i)} {\bm \theta} _{k_l'(j)}^{\rm H} {\bm \theta} _{k_l'(j)} {\bm \theta} ^{\rm H}_{k_h(i)} \right\} {\bf I}_{MN},
\label{PZF_H_2_1}
\end{align} }

\noindent where $(a)$ in~\eqref{PZF_H_2_1} is achieved by substituting~\eqref{MIMOH} and applying the properties of Kronecker product, and $(b)$ is due to the path independence.
Here, we also need to obtain $\mathbb E \left\{ {\bm \theta} _{k_h(i)} {\bm \theta} _{k_l'(j)}^{\rm H} {\bm \theta} _{k_l'(j)} {\bm \theta} ^{\rm H}_{k_h(i)} \right\}$ which is expressed as $\mathbb E \left\{ {\bm \theta} _{k_h(i)} \mathbb E \left\{{\bm \theta} _{k_l'(j)}^{\rm H} {\bm \theta} _{k_l'(j)}\right\} {\bm \theta} ^{\rm H}_{k_h(i)} \right\}$. Then, we have
\begin {align}
 &\mathbb E \left\{ {\bm \theta} _{k_l'(j)}^{\rm H}   {\bm \theta} _{k_l'(j)} \right\} \notag \\
&={\left[\!\! {\begin{array}{*{20}{c}}
      \mathbb E \{w_{(j)}^0\}         & \mathbb E \{w_{(j)}^1\}           & \ldots & \mathbb E \{w_{(j)}^{N_t-1}\}    \\
      \mathbb E \{w_{(j)}^{-1}\}      & \mathbb E \{w_{(j)}^0\}           & \ldots & \mathbb E \{w_{(j)}^{N_t-2}\}  \\
      \vdots             & \vdots                & \ddots & \vdots\\
      \mathbb E \{w_{(j)}^{-N_t+1}\}  & \mathbb E \{w_{(j)}^{-N_t+2}\}    & \ldots & \mathbb E \{w_{(j)}^0\}
      \end{array}} \right]} ,
      \vspace{1em}
\end {align}
\noindent where $w_{(j)}={\rm exp}(-j\pi \sin \phi_{(j)} )$. We assume $\sin \phi_{(j)}$ is a random variable with equal probability in the range of $[-1, 1]$. Hence, with $n=1,\ldots, N_t-1$, we have
\vspace{-0.2em}
\begin{align}
\mathbb E \{w_{(j)}^n\}
&\stackrel{(a)}{=} \int_{-\infty}^{\infty} {\rm exp}(-j n \pi x ) { p}(\sin \phi_{(j)}=x) \,dx \notag\\    
&=\frac{{\rm exp}(-j n \pi  ) -{\rm exp}(j n \pi  )}{-2jn\pi} \notag\\
&\stackrel{(b)}{=}\frac{2j \sin( -n \pi ) }{-2jn\pi} 
=\sinc(n\pi) 
=0,
\vspace{1em}
\label{Proof_angle}
\end{align}
where $(a)$ in~\eqref{Proof_angle} is due to the law of total expectation, and $(b)$ is derived using Euler's formula. 
Based on~\eqref{Proof_angle}, we have 
\begin{align}
    \mathbb E \left\{ {\bm \theta} _{k_l'(j)}^{\rm H}   {\bm \theta} _{k_l'(j)} \right\}={\bf I}_{N_t}.
    \vspace{1em}
\end{align}
Therefore, 
\begin{align}
  \mathbb E \left\{ {\bm \theta} _{k_h(i)} \mathbb E \left\{{\bm \theta} _{k_l'(j)}^{\rm H} {\bm \theta} _{k_l'(j)}\right\} {\bm \theta} ^{\rm H}_{k_h(i)} \right\}&=
\mathbb E \left\{ {\bm \theta} _{k_h(i)} {\bm \theta} ^{\rm H}_{k_h(i)} \right\} \nonumber\\
&={\bf I}_{N_t},
\end{align}
where the last equality was obtained by following similar steps as in~\eqref{Proof_angle}. Therefore,~\eqref{PZF_H_2_1} is simplified to
\begin{align}
\mathbb E \left\{\Htd _{k_h}   (\Htd _{k_l'})^{\rm H}   \Htd _{k_l'} (\Htd _{k_h})^{\rm H}\right\} 
={\beta_{k_h}\beta_{k'_l}} N_t {\bf I}_{MN}.
\vspace{1em}
\end{align} 
Accordingly,~\eqref{PZF_H_2} can be approximated as
\begin{align}
\mathbb E \left\{ \D_{k_h k_l'} \D_{k_h k_l'}^{\rm H}  \right\} 
&\approx (\alpha^{\MRT}_{k'_l}) ^2\rho\eta_{k_l'} \beta_{k_h}\beta_{k'_l} N_t {\bf I}_{MN}.
\label {PZF_H_3}
\vspace{0.3em}
\end{align} 

\begin{Corollary}
The achievable SE with PZF precoding for the HM-UE in~\eqref{PZF_H_1} can be further approximated as
 \begin{align}
{\rm{SE}}_{k_h}^\PZF
&\!\approx\!\alpha_{\rm SE} MN {\log _2} \Bigg( 1+ \frac{\alpha_\PZF^2\rho\eta_{k_h}} { 1 \!+\! {\sum_{k_l'\in \Kl} (\alpha^{\MRT}_{k'_l}) ^2 \beta_{k_h}\beta_{k'_l} \rho\eta_{k_l'}N_t}    }  \Bigg) .
\label{PZF_app_H}
\vspace{1em}
\end{align} 
\end{Corollary}

\begin{proposition}~\label{theorem:MRT}
  With PZF precoding and HL grouping, the SE for the $k_l$-th LM-UE is shown in~\eqref{SE_PZF_2} at the top of the next page.
\begin{figure*}
\begin{align}
{\rm{SE}}_{k_l}^\MRT&=\alpha_{\rm{SE}} {\log _2}\det \Bigg( {\bf I}_{L_dN}+ \frac{\beta _{k_l} N_t \rho\eta_{k_l}{\bf I}_{L_dN}}   {{\bf I}_{L_dN} +  \sum_{k'_h\in \Kh}  \mathbb E \big\{ \D_{k_l k_h'} \D_{k_l k_h'}^{\rm H}  \big\} + \sum_{k'_l\in \Kl} \mathbb E \big\{ \D_{k_l k_l'} \D_{k_l k_l'}^{\rm H}  \big\}  - {\beta _{k_l} N_t \rho\eta_{k_l}} {\bf I}_{L_dN} }  \Bigg),
 \label{SE_PZF_2}
\end{align}
{\text{with}}
\begin{align}
\mathbb E \big\{ \D_{k_l k_h'} \D_{k_l k_h'}^{\rm H}  \big\}
&={\rho\eta_{k_h'}}  \left(  {\bf I}_{N}\otimes{{\bf{F}}_{L_{d}}}{{\bf R}_{\rm CP}}  \right)  \mathbb E \big\{{{\qH}}^{\rm TD}_{k_l} {\bf W}_{k'_h}^\PZF ({\bf W}_{k'_h}^\PZF)^{\rm H} ({{\qH}}^{\rm TD}_{k_l})^{\rm H}\big\} \left(  {\bf I}_{N}\otimes{{\bf R}_{\rm CP}^{\rm H}}{{\bf{F}}_{L_{d}}}  \right).
\label{NMSE_1}\\
\mathbb E \big\{ \D_{k_l k_l'} \D_{k_l k_l'}^{\rm H}  \big\} 
&={\rho\eta_{k_l'}} \left(  {\bf I}_{N}\otimes{{\bf{F}}_{L_{d}}}{{\bf R}_{\rm CP}}  \right) \mathbb E \left\{{{\qH}}^{\rm TD}_{k_l} {\bf W}_{k'_l}^\MRT \left (  {\bf I}_{N}\otimes{{\bf A}_{\rm CP} {\bf A}_{\rm CP}^{\rm H}}  \right )
({\bf W}_{k'_l}^\MRT)^{\rm H} ({{\qH}}^{\rm TD}_{k_l})^{\rm H}\right\} \left(  {\bf I}_{N}\otimes{{\bf R}_{\rm CP}^{\rm H}}{{\bf{F}}_{L_{d}}}  \right) . 
\end{align} 
\rule{\textwidth}{0.3mm}
\end{figure*} 
\end{proposition}
\begin{proof}
See Appendix~\ref{APX:4}.    
\end{proof}

According to Propositions 2 and 3, with the help of the PZF, intra-group interference for HM-UEs can be eliminated. Yet, HM-UEs still suffer from inter-group interference, while LM-UEs will experience both intra-group and inter-group interference. To manage the interference, in the next section, we will propose two power allocation schemes. For the sake of simplifying the power allocation design, we then make some approximations for the LM-UEs. 

To further simplify $\mathbb E \big\{ \D_{k_l k_l'} \D_{k_l k_l'}^{\rm H}  \big\} $, similar as in~\eqref{PZF_H_1} and~\eqref{PZF_H_2_1}, 
for the intra-group interference from user $k_l'$, 
where $k_l'\in \Kl$, $k_l'\neq k_l$, by applying ${\bf A}_{\rm CP} {\bf A}_{\rm CP}^{\rm H} \approx {\bf I}_M$, we have
\begin{align}
&\mathbb E \big\{ \D_{k_l k_l'} \D_{k_l k_l'}^{\rm H}  \big\} \notag\\
&\approx {\rho\eta_{k_l'}} \left(  {\bf I}_{N}\otimes{{\bf{F}}_{L_{d}}}{{\bf R}_{\rm CP}}  \right) \mathbb E \left\{{{\qH}}^{\rm TD}_{k_l} {\bf W}_{k'_l}^\MRT ({\bf W}_{k'_l}^\MRT)^{\rm H} ({{\qH}}^{\rm TD}_{k_l})^{\rm H}\right\} \notag\\
&\hspace{4em}
\times\left(  {\bf I}_{N}\otimes{{\bf R}_{\rm CP}^{\rm H}}{{\bf{F}}_{L_{d}}}  \right) \notag\\
&= \frac{ (\alpha^{\MRT}_{k'_l}) ^2 {\rho\eta_{k_l'}}\beta_{k_l}\beta_{k'_l}}{P^2} \sum_{i=1}^{P} \sum_{j=1}^{P} \mathbb E \left\{ \big\vert {\bm \theta} _{k_l(i)} {\bm \theta} _{k_l'(j)}^{\rm H} \big\vert^2 \right\} {\bf I}_{L_dN}\notag\\
&=  (\alpha^{\MRT}_{k'_l}) ^2 {\rho\eta_{k_l'}} \beta_{k_l}\beta_{k'_l} N_t {\bf I}_{L_dN}. 
\end{align}
With $k_l'= k_l$, we have
\begin{align}
&\mathbb E \big\{ \D_{k_l k_l} \D_{k_l k_l}^{\rm H}  \big\} 
\approx {\rho\eta_{k_l}} (\alpha^{\MRT}_{k_l}) ^2 \left(  {\bf I}_{N}\otimes{{\bf{F}}_{L_{d}}}{{\bf R}_{\rm CP}}  \right) \notag \\
&\hspace{0em}\times\mathbb E \left\{{{\qH}}^{\rm TD}_{k_l} ({{\qH}}^{\rm TD}_{k_l})^{\rm H} {{\qH}}^{\rm TD}_{k_l} ({{\qH}}^{\rm TD}_{k_l})^{\rm H}\right\} \left(  {\bf I}_{N}\otimes{{\bf R}_{\rm CP}^{\rm H}}{{\bf{F}}_{L_{d}}}  \right) .
\label{DD_ll_1}
\end{align} 
Note that, based on~\eqref{MIMOH}, we have
\begin{align}
 &\mathbb E \left\{{{\qH}}^{\rm TD}_{k_l} ({{\qH}}^{\rm TD}_{k_l})^{\rm H} {{\qH}}^{\rm TD}_{k_l} ({{\qH}}^{\rm TD}_{k_l})^{\rm H}\right\} \notag\\
 &=\beta _{k_l}^2 \sum_{i=1}^{P} \sum_{j=1}^{P}\sum_{m=1}^{P}\sum_{n=1}^{P} \!\mathbb E \left\{ {\bf{H}}^{\rm{TD}}_{k_l (i)}  ({\bf{H}}^{\rm{TD}}_{k_l (j)})^{\rm H}  {\bf{H}}^{\rm{TD}}_{k_l (m)}  ({\bf{H}}^{\rm{TD}}_{k_l (n)})^{\rm H}\right\} \notag\\
 &\stackrel{(a)}{=}\beta _{k_l}^2 \sum_{i=1}^{P}  \mathbb E \left\{ {\bf{H}}^{\rm{TD}}_{k_l (i)}  ({\bf{H}}^{\rm{TD}}_{k_l (i)})^{\rm H}  {\bf{H}}^{\rm{TD}}_{k_l (i)}  ({\bf{H}}^{\rm{TD}}_{k_l (i)})^{\rm H}\right\} \notag \\
&~~ +\beta _{k_l}^2 \sum_{i=1}^{P} \sum_{j=1, j\neq i}^{P} \mathbb E \left\{ {\bf{H}}^{\rm{TD}}_{k_l (i)}  ({\bf{H}}^{\rm{TD}}_{k_l (j)})^{\rm H}  {\bf{H}}^{\rm{TD}}_{k_l (j)}  ({\bf{H}}^{\rm{TD}}_{k_l (i)})^{\rm H}\right\} \notag \\
&~~+\beta _{k_l}^2 \sum_{i=1}^{P} \sum_{j=1, j\neq i}^{P} \mathbb E \left\{ {\bf{H}}^{\rm{TD}}_{k_l (i)}  ({\bf{H}}^{\rm{TD}}_{k_l (i)})^{\rm H}  {\bf{H}}^{\rm{TD}}_{k_l (j)}  ({\bf{H}}^{\rm{TD}}_{k_l (j)})^{\rm H}\right\} \notag \\
&\stackrel{(b)}{=}\beta _{k_l}^2 \left ( \frac{2}{P}N_t^2 + \frac{P^2-P}{P^2} N_t + \frac{P^2-P}{P^2} N_t^2 \right ) {\bf I}_{MN}\notag \\
&=\beta _{k_l}^2 N_t\left ( N_t+1+\frac{N_t-1}{P} \right ) {\bf I}_{MN},
\label{E_llll}
\end{align} 
where $(a)$ in~\eqref{E_llll} is based on the independence of different paths, and the detailed proof of $(b)$ will be provided in Appendix~\ref{APX:3}. Therefore,~\eqref{DD_ll_1} can be further simplified as
\begin{align}
\mathbb E \big\{ \D_{k_l k_l} \D_{k_l k_l}^{\rm H}  \big\} 
&\approx   {\rho\eta_{k_l}} \big(\alpha^{\MRT}_{k_l}\big) ^2 \beta _{k_l}^2
\nonumber\\
&\hspace{2em}
\times N_t \left (N_t+1+\frac{N_t-1}{P} \right) {\bf I}_{L_dN},
\end{align} 
and $\mathbb E \big\{ \D_{k_l k_h'} \D_{k_l k_h'}^{\rm H}  \big\}$ is represented as in~\eqref{48} at the top of the next page.
\begin{figure}
    \centering
    \vspace{-1.5em}
    \includegraphics[width=0.9\linewidth]{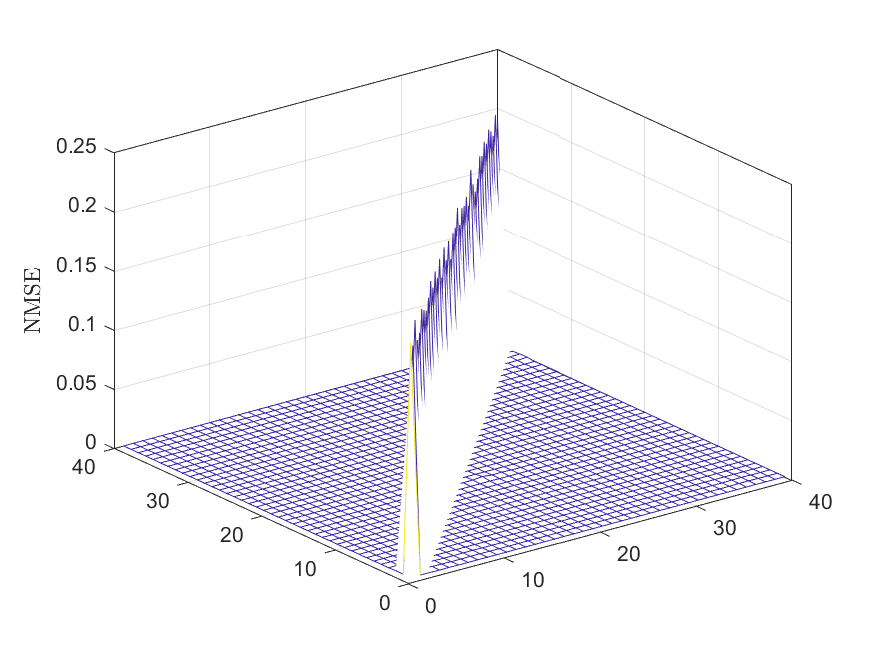}
    \caption{NMSE as in~\eqref{NMSE}.}
    \label{NMSE_fig}
\end{figure}

\begin{figure*}
\begin{align}
\mathbb E \big\{ \D_{k_l k_h'} \D_{k_l k_h'}^{\rm H}  \big\}
&={\rho\eta_{k_h'}}  \left(  {\bf I}_{N}\otimes{{\bf{F}}_{L_{d}}}{{\bf R}_{\rm CP}}  \right)  \mathbb E \big\{{{\qH}}^{\rm TD}_{k_l} {\bf W}_{k'_h}^\PZF ({\bf W}_{k'_h}^\PZF)^{\rm H} ({{\qH}}^{\rm TD}_{k_l})^{\rm H}\big\} \left(  {\bf I}_{N}\otimes{{\bf R}_{\rm CP}^{\rm H}}{{\bf{F}}_{L_{d}}}  \right) \notag\\
&=(\alpha^{\PZF}_{k_h})^2 {\rho\eta_{k_h'}}  \left(  {\bf I}_{N}\otimes{{\bf{F}}_{L_{d}}}{{\bf R}_{\rm CP}}  \right)  \mathbb E \big\{{{\qH}}^{\rm TD}_{k_l} (\mH^\PZF)^{\rm H} \left (  \mH^\PZF (\mH^\PZF)^{\rm H} \right )^{\!-1} \notag\\ 
&\hspace{2em}\times\!\left ( \!{\qb_{K_h}^{(k_h')}} ({\qb_{K_h}^{(k_h')}})^{\rm H}\! \otimes \! \qI_{MN} \!\right)  \left (  \mH^\PZF (\mH^\PZF)^{\rm H} \!\right )^{-\! 1}\mH^\PZF ({{\qH}}^{\rm TD}_{k_l})^{\rm H}\big\} \left(  {\bf I}_{N}\otimes{{\bf R}_{\rm CP}^{\rm H}}{{\bf{F}}_{L_{d}}}  \right).
\label{48}
\end{align} 
\rule{\textwidth}{0.1mm}
\end{figure*}
Note that the small-scale channel coefficients of different users are independent from each other. Furthermore, the matrix inversion does not affect this independence. Therefore, we have the approximation 
\begin{align} 
\mathbb E \big\{ \D_{k_l k_h'} \D_{k_l k_h'}^{\rm H}  \big\}
\approx {\rho\eta_{k_h'}} \beta_{k_l} {\bf I}_{L_d N}.
\label{NMSE_2}
\end{align}
\noindent To verify the tightness of this approximation, we first define the normalized mean square error (NMSE) as
\begin{align}
    {\rm {NMSE}}=\frac{|\mathbb E \big\{ \D_{k_l k_h'} \D_{k_l k_h'}^{\rm H}  \big\}- {\rho\eta_{k_h'}} \beta_{k_l} {\bf I}_{L_d N}|^2} {|\mathbb E \big\{ \D_{k_l k_h'} \D_{k_l k_h'}^{\rm H}  \big\}|^2}.
    \label{NMSE}
\end{align}
The numerical result for the NMSE matrix of size $L_d N\times L_d N$ is illustrated in Fig~\ref{NMSE_fig}. From the simulation results, we notice that~\eqref{NMSE_2} gives a close approximation to~\eqref{NMSE_1}.

\begin{Corollary}
The achievable SE with PZF precoding for the $k_l$-th LM-UE in~\eqref{SE_PZF_2} can be further approximated as
\begin{align}
{\rm{SE}}_{k_l}^\MRT
&\approx \alpha_{\rm SE} {L_d}N {\log _2} \bigg(1+\frac{{\beta _{k_l} N_t \rho\eta_{k_l}}} {  1+ { \Psi_{k_l}} -\beta_{k_l}N_t \rho \eta_{k_l} }  \bigg),
 \label{PZF_app_L}
\end{align} 
with 
\begin{align}
{ \Psi_{k_l}}
&\triangleq \sum_{k'_h\in \Kh} \rho \eta_{k'_h} \beta_{k_l} 
+ {\rho\eta_{k_l}} (\alpha^{\MRT}_{k_l}) ^2 \beta _{k_l}^2 N_t \left ( N_t+1+\frac{N_t-1}{P} \right ) \notag \\
&\hspace{2em}+ \sum\nolimits_{k'_l\in \Kl, k'_l\neq k_l} (\alpha^{\MRT}_{k'_l}) ^2 {\rho\eta_{k_l'}} \beta_{k_l}\beta_{k'_l} N_t .
\end{align} 
\end{Corollary}
Note that the tightness of our approximations is verified later in the numerical results section via simulations. 

\vspace{-1em}
\subsection{Complexity Analysis}
According to the detailed precoding design for FZF and PZF precoding, we then investigate the complexity of both schemes in terms of the big $O$ function. 

Firstly, based on the FZF precoding design shown in~\eqref{FZF} and~\eqref{FZF_L}, the main computational complexity comes from the calculation of the inversion of the matrix ${\mH}^{\rm FZF} ({\mH}^{\rm FZF})^{\rm H} \in {\mathbb{C}}^{KMN\times KMN}$. Therefore, the complexity for FZF can be approximated by $O((KMN)^3)$ for both HM and LM-UEs.

By focusing on the PZF precoding scheme, we see that for the HM-UEs precoding design in~\eqref{PZF}, the main complexity comes from the inversion of the matrix $\mH^\PZF (\mH^\PZF)^{\rm H} \in {\mathbb{C}}^{K_hMN\times K_hMN}$. The complexity can then be approximated by $O((K_hMN)^3)$. 
For the MRT of the LM-UEs shown in~\eqref{W_MRT}, the Hermitian of a matrix is a linear operation. The main complexity of the precoding can be represented by $O(MN\times N_tMN)=O(N_t(MN)^2)$.

\begin{table*}
    \centering
        \caption{Complexity of the considered schemes.}
    \begin{tabular}{|c|c|c|c|}
    \hline
        HM-UEs with FZF   & LM-UEs with FZF  & HM-UEs with PZF   & LM-UEs with PZF\\
          \hline
        $O(KMN)^3$ & $O(KMN)^3$ & $O(K_h MN)^3$ & $O(N_t(MN)^2)$ \\
        \hline
    \end{tabular}
    \label{complexity}
\end{table*}

The complexity of different precoding schemes is summarized in Table~\ref{complexity}.
We can see that in Table~\ref{complexity}, the complexity for different users is in descending order from left to right. By further considering the hardware requirement for OTFS, HM-UEs with FZF have the highest complexity, while the LM-UEs with PZF have the lowest complexity.
Moreover, we notice that the complexity for FZF and PZF depends on the total number of users $K$ and the number of HM-UEs $K_h$, respectively. 
This suggests that PZF generally has a lower complexity than FZF. In the case of $K=K_h$, FZF ends up with the same complexity as PZF. Therefore, FZF can be seen as a special case of the PZF with all users identified as HM-UEs. 
\vspace{-0.5em}
\section{Power Allocation}~\label{sec:power}
In this section, we introduce two power allocation schemes based on the derived closed-form SE at the BS to ensure fairness in the system. 
Note that based on statistical CSI, the derived closed-form SE expressions offer a significant reduction in complexity and overhead required for power allocation.
First, we explore the max-min fairness power control scheme. The power allocation coefficients $\eta_k$, $k=1,\ldots, K$ are computed at the BS based on the given realization of large-scale fading. With max-min power control, we determine the power allocation coefficients that maximize the minimum SE among all users. 
The max-min fairness power allocation optimization problem can be formulated as follows
\!
\begin{align}
\max_{\{\eta_k\}} \quad& \min_{k=1,\ldots,K}  {\rm{SE}}_{k} \notag\\
\mbox{subject to} \quad& \sum\nolimits_{k=1}^{K} \eta_k \leq 1 \notag\\
   & 0 \le \eta_k, k=1,\ldots,K.
\label{PC_0}
\end{align}

Next, we consider a weighted max-min power control design. By inspecting Propositions 2 and 3, we observe that LM-UEs experience more interference than HM-UEs, resulting in an overall lower SE. Hence, the max-min power control design in~\eqref{PC_0} will undermine the SE for HM-UEs. To address this issue, instead of providing fairness to all users, we  consider a proportional fairness maximization, which is formulated as 
\begin{align}
\max_{\{\eta_k\}} \hspace{0.5em}& \Big\{ \alpha_w w_h \min_{k_h \in \mathcal{K}_h}  {\rm{SE}}_{k_h} + \alpha_w w_l \min_{k_l\in \mathcal{K}_l}  {\rm{SE}}_{k_l}\Big\}\notag\\
\mathrm{s.t.} \,\,
		\hspace{0.5em}&\sum\nolimits_{k=1}^{K} \eta_k \leq 1 \notag\\
   & 0 \le \eta_k, k=1,\ldots,K,
\label{PC_1}
\end{align}
where $w_h$ and $w_l$ are the weighting coefficients for HM and LM-UEs, respectively. Moreover, $\alpha_w\triangleq \frac{1}{w_h+w_l}$ is the normalization weighting coefficient. 

\vspace{-0.8em}
\subsection{FZF Precoding}
With FZF precoding design, we consider the max-min power allocation design. By invoking~\eqref{SE_HFZF} and~\eqref{SE_LFZF}, and noticing that a logarithm function is monotonically increasing,~\eqref{PC_0} is equivalently reformulated as
\!
\begin{align}
\max_{\{\eta_k\}} \quad& \min_{k=1,\cdots,K}  
		(1+\alpha_{\FZF}^{2} \rho\eta_{k})^{\alpha_ o}  \notag\\
\mbox{s.t.} \quad & \sum\nolimits_{k=1}^{K} \eta_k \leq 1 \notag\\
   & 0 \le \eta_k, k=1,\ldots,K.
   \label{PC_FZF}
\end{align}

\begin{algorithm}
\caption{Bisection algorithm for solving~\eqref{PC_PZF}}\label{alg:2}
\begin{itemize}
   \item[(1)] {\emph{Initialization}}: Choose the initial values of $t_{\rm max}$ and $t_{\rm min}$, where $t_{\rm max}$ and $t_{\rm min}$ define a range of objective function values. Set tolerance $\epsilon >0$.
   \item[(2)] Set $t:=\frac{t_{\rm max}+t_{\rm min}}{2}$ and solve the following convex feasibility problem:
   \begin{align}
   \begin{cases}
       & t \leq {\rm{SE}}_{k} , \quad k=1,\cdots,K \notag\\
       & \sum_{k=1}^{K} \eta_k \leq 1 \notag\\
       & 0 \le \eta_k, \quad k=1,\cdots,K
   \end{cases}
   \end{align}
   \item[(3)] If the problem in Step $2$ is feasible, set $t_{\min}:=t$; else set $t_{\rm max}:=t$.
   \item[(5)] Stop if $t_{\rm max}-t_{\rm min}< \epsilon$. Otherwise, go to Step $2$.
\end{itemize}
\label{a_2}
\end{algorithm}
\setlength{\textfloatsep}{0.1cm}

Note that in~\eqref{PC_FZF}, we approximate $\alpha_o=1$ for $k\in \Kh$ and $\alpha_o=\frac{L_d}{M}$ for $k\in\Kl$.
As the optimization problem~\eqref{PC_FZF} is a quasiconvex problem on the non-negative interval, the optimization problem can be efficiently solved using CVX~\cite{cvx}. 

According to~\cite{Tam2016TWC}, the computational complexity to solve the feasibility problem~\eqref{PC_FZF} is $\mathcal{O}\big( {\sqrt{n_l\!+n_q}} (n_{l} \!+ n_{v} \!+ n_q) n_v^2 \big)$, where $n_{l} \!=\!K\!+\!K_h\!+\!1$ denotes the number of linear constraints, $n_v \!=\!K$ is the number of real-valued scalar decision variables, and $n_q\!=\!K_l$ is the number of quadratic constraint.

\vspace{-0.9em}
\subsection{PZF Precoding}
\subsubsection{Max-min power control}
For PZF precoding, we first consider the max-min power control for all the users. To this end, we use the SE for HM- and LM-UEs provided in~\eqref{PZF_app_H} and~\eqref{PZF_app_L}, respectively. Therefore, by introducing the auxiliary variable $t$, problem~\eqref{PC_0} is equivalent to
\vspace{-0.2em}
\begin{align}
\max_{\{\eta_k\},t}\quad &  t \notag\\
\mbox{s.t.} \quad
& t \leq {\rm{SE}}_{k} , \quad k=1,\ldots,K \notag\\
& \sum\nolimits_{k=1}^{K} \eta_k \leq 1 \notag\\
& \eta_k\ge 0, \quad k=1,\ldots,K.
\label{PC_PZF}
\end{align}
Based on~\eqref{PZF_app_H} and~\eqref{PZF_app_L}, for a given $t$, all the inequalities involved in~\eqref{PC_PZF} are linear. Hence,~\eqref{PC_PZF} is a quasi-linear problem and can be solved by using the bisection technique and solving linear feasibility problems~\cite{Boyd}. Specifically,\textbf{ Algorithm~\ref{a_2}} solves~\eqref{PC_PZF}. 

According to~\cite{Tam2016TWC}, the  per-iteration cost to solve the feasibility problem~\eqref{PC_PZF} is $\mathcal{O}\big( (n_{l} + n_{v}) n_v^2n_{l}^{0.5}\big)$, where $n_{l} \!=\!2K+1 $ 
and $n_v =K+1$. 
Therefore, the overall complexity of the bisection algorithm is $\lceil\log2((t_{\max}-t_{\min})/\epsilon)\rceil\mathcal{O}\big( (n_{l} + n_{v}) n_v^2n_{l}^{0.5}\big)$.

\subsubsection{Weighted max-min power control}
Due to the different interference levels for HM and LM-UEs with PZF precoding, considering max-min power allocation will result in an overall much lower SEs for all users. Therefore, we propose a weighted max-min power allocation scheme, where performance fairness is promoted for each group.
To enable this, we recast the optimization problem~\eqref{PC_1} as 
\vspace{-0.2em}
\begin{subequations}\label{eq:opt:P2}
\begin{align}
&\max_{\{\eta_k,t_h,t_l,T_h,T_l \}} \hspace{-0.2em}  \alpha_w w_h t_h + \alpha_w w_l t_l  \label{56a}\\
& ~~~~~~~ \mathrm{s.t.} \,\,\,
T_h\!\leq\! \frac{\alpha_\PZF^2\rho\eta_{k_h}} { 1 \!+\! N_t \rho\beta_{k_h}{\sum^{K_l}_{k_l'=1} (\alpha^{\MRT}_{k'_l}) ^2 \beta_{k'_l} \eta_{k_l'}}},~ \forall k_h\!\in \!\Kh \label{56b} \\
&\hspace{4.3em} T_l\leq \frac{{\beta _{k_l} N_t \rho\eta_{k_l}}} {1+ {\Psi_{k_l}} -\beta_{k_l}N_t \rho \eta_{k_l} }, ~ \forall k_l\in \Kl \label{56c}\\
&\hspace{4.3em} 2^{t_h}-1 \le T_h \\
&\hspace{4.3em} 2^{\frac{t_l}{\alpha_{\rm SE}}}-1 \le T_l \\
&\hspace{4.3em} \sum\nolimits_{k=1}^{K} \eta_k \leq 1 \\
&\hspace{4.3em} 0 \le \eta_k,\quad k=1,\ldots,K.
\label{PC_1_2}
\end{align}
\end{subequations} 

Problem~\eqref{eq:opt:P2} is difficult to solve due to the non-convex constraints~\eqref{56b} and~\eqref{56c}. To deal with these constraints, we first express~\eqref{56b} and~\eqref{56c} as
\begin{align}
    T_h + {\sum_{k_l'\in \Kl} (\alpha^{\MRT}_{k'_l}) ^2 \beta_{k_h}\beta_{k'_l} N_t \rho \eta_{k_l'} T_h} 
    \leq {\alpha_\PZF^2\rho\eta_{k_h}},
    \label{59}
\end{align}
and
\begin{align}
    T_l +  { \Psi_{k_l}} T_l -\beta_{k_l}N_t \rho  \eta_{k_l} T_l 
    \leq {\beta _{k_l} N_t \rho\eta_{k_l}},
    \label{60}
\end{align}
respectively. We notice that the non-convexity in~\eqref{59} and~\eqref{60} is due to the product terms $\eta_{k_l'} T_h$ and $\eta_{k_l} T_l$.  To deal with this challenge, we further consider the application of the successive convex approximation techniques. We apply the following upper-bound 
\begin{align}~\label{eq:bound}
   xy\leq \frac{1}{4} [(x+y)^2-2(x_{(n)}-y_{(n)})(x-y)+(x_{(n)}-y_{(n)})^2] ,
\end{align}
for non-negative variables $x$ and $y$, where $x_{(n)}$ and $y_{(n)}$ denote the approximation values for $x$ and $y$ for the $n$-th iteration, respectively. 
To obtain the values of $x_{(n)}$ and $y_{(n)}$ in each case, we first assign an initial value for $x_{(n)}$ and $y_{(n)}$, respectively, and update $x_{(n)}$ and $y_{(n)}$ with the calculated $x$ and $y$ values at the end of each iteration. 
The iteration ends when $|x_{(n)}-x|$ and $|y_{(n)}-y|$ is less than a certain threshold or the number of iterations reaches the iteration threshold.
With the help of~\eqref{eq:bound}, for $T\in [T_h, T_l] $, we first define
\begin{align}
    C({{\eta_{k'} },{T} })
    &\triangleq 
    \frac{1}{4} [(\eta_{k'}+T)^2-2({\eta_{k'}}_{(n)}-{T}_{(n)})(\eta_{k'}-{T}) \notag \\
    &~~+({\eta_{k'}}_{(n)}-{T}_{(n)})^2] .
\end{align}
Hence, we can express~\eqref{59} and~\eqref{60} as  
\begin{align}
    \sum_{k_l'\in \Kl} (\alpha^{\MRT}_{k'_l}) ^2 \beta_{k_h}\beta_{k'_l} N_t \rho  C ({{\eta_{k'_l} },{T_h} })
    \leq {\alpha_\PZF^2\rho\eta_{k_h}} -T_h ,
\end{align}
\begin{align}
\tilde{{ \Psi}} _{k_l}
-\beta_{k_l}N_t \rho  
C({{\eta_{k_l} },{T_l} })
\leq {\beta _{k_l} N_t \rho\eta_{k_l}}- T_l ,
\label{61}
\end{align}
where 
\begin{align}
\tilde{{ \Psi}}_{k_l}
&=\!\!\!\sum_{k'_h\in \Kh}\!\! \rho  \beta_{k_l} C({{\eta_{k'_h} },{T_l} }) \!+\!{\rho} (\alpha^{\MRT}_{k_l}) ^2 \beta _{k_l}^2 N_t  \bigg(  N_t\!+\!1\!+\!{ \frac{N_t-1}{P}} \bigg) \notag\\
&~~ \times C({{\eta_{k_l} },{T_l} })+\!\!\!\!\sum_{k'_l\in \Kl, k'_l\neq k_l} \!\!(\alpha^{\MRT}_{k'_l}) ^2 {\rho} \beta_{k_l}\beta_{k'_l} N_t C({{\eta_{k'_l} },{T_l} }).
\end{align}

Since the left-hand side of the ~\eqref{61} is still non-convex due to the presence of concave function, we further apply the inequality $x^2 \ge x_{(n)}(2x-x_{(n)})$ as following
\begin{align}
 C'({{\eta_{k_l} },{T_l} }) 
 &  \ge C({{\eta_{k_l} },{T_l} })  \notag\\
 &\triangleq  \frac{1}{4} \bigg[\big(\eta_{k_l(n)}\!+\!{T_{l(n)}}\big)\! \big (2(\eta_{k_l}\!+\!{T_{l}})\!-\!(\eta_{k_l(n)}\!+\!{T_{l(n)}}) \big ) \notag\\
 &~~-2\big({\eta_{k_l}}_{(n)}\!-\!{T}_{l(n)}\big)\big(\eta_{k_l}\!-\!{T}_l\big)\!+\!\big({\eta_{k_l}}_{(n)}\!-\!{T}_{l(n)}\big)^2\bigg].
\end{align}
Thus, the optimization problem~\eqref{PC_1} can be expressed as
  \begin{align}
\max_{\{\eta_k,T_h,T_l,t_h,t_l,t\}} \hspace{0.5em}& t  \notag\\
\mathrm{s.t.} \,\,
		\hspace{2em}&
\alpha_w w_h t_h  + \alpha_w w_l t_l \geq t \notag\\
& \sum_{k_l'\in \Kl} (\alpha^{\MRT}_{k'_l}) ^2 \beta_{k_h}\beta_{k'_l} N_t \rho  C ({{\eta_{k_l} },{T_h} }) 
\notag\\
&\hspace{3em}\leq {\alpha_\PZF^2\rho\eta_{k_h}} -T_h , ~ k_h\in \Kh  \notag\\
& \tilde{{ \Psi}}_{k_l} 
-\beta_{k_l}N_t \rho  C'({{\eta_{k_l} },{T_l} })
\notag\\
&\hspace{3em}
\leq {\beta _{k_l} N_t \rho\eta_{k_l}}- T_l , ~ k_l\in \Kl  \notag\\
& T_h \ge 2^{t_h}-1\notag\\
& T_l \ge 2^{\frac{t_l}{\alpha_{\rm SE}}}-1\notag\\
& \sum_{k=1}^{K} \eta_k \leq 1 \notag\\
& \eta_k \ge 0,\quad k=1,\ldots,K.
\label{PC_22}
\end{align}  

As the optimization problem in~\eqref{PC_22} is a convex problem, again, we can solve it directly using the bisection technique and solving linear feasibility problems, as shown in \textbf{Algorithm~\ref{a_1}}.

According to~\cite{Tam2016TWC}, the overall complexity of the bisection algorithm is $\lceil\log2((t_{\max}-t_{\min})/\epsilon)\rceil\mathcal{O}\big( (n_{l} + n_{v}) n_v^2n_{l}^{0.5}\big)$, where $n_{l} =2K+4 $ denotes the number of linear constraints and $n_v =K+5$ is the number of real valued scalar decision variables.

\begin{algorithm}[t]
\caption{Bisection algorithm for solving~\eqref{PC_22}}\label{alg:cap}
\begin{itemize}
   \item[(1)] {\emph{Initialization}}: Choose the initial values of $t_{\rm max}$ and $t_{\rm min}$, where $t_{\rm max}$ and $t_{\rm min}$ define a range of objective function values. Set tolerance $\epsilon_1, \epsilon_2 >0$, iteration number $n_i=0$, and the initial value for $T_{h(n)}$, $T_{l(n)}$ and ${\boldsymbol{\eta_{(n)}}}\triangleq [\eta_{1},\ldots,\eta_{K_h},\eta_{1},\ldots,\eta_{K_l}]$.
   \item[(2)] Set $t:=\frac{t_{\rm max}+t_{\rm min}}{2}$, $n_i:=n_i+1$, and solve the following convex feasibility problem:
   {\small
   \begin{align}
   \begin{cases}
       & \alpha_w w_h t_h  + \alpha_w w_l t_l \geq t \notag\\
       & \sum_{k'_l\in \Kl} (\alpha^{\MRT}_{k'_l}) ^2 \beta_{k_h}\beta_{k'_l} N_t \rho  C ({{\eta_{k_l} },{T_h} }) \leq {\alpha_\PZF^2\rho\eta_{k_h}} -T_h , \notag\\
        &\hspace{15em} k_h\in \Kh  \notag\\
       & \tilde{{\Psi}}_{k_l} -\beta_{k_l}N_t \rho  C'({{\eta_{k_l} },{T_l} })\leq {\beta _{k_l} N_t \rho\eta_{k_l}}- T_l , ~ k_l\in \Kl  \notag\\
       & T_h \ge 2^{t_h}-1\notag\\
       & T_l \ge 2^{\frac{t_l}{\alpha_{\rm SE}}}-1\notag\\
       & \sum_{k=1}^{K} \eta_k \leq 1 \notag\\
       & \eta_k \ge 0,\quad k=1,\cdots,K.
   \end{cases}
   \end{align}}
   \item[(3)] If the problem is feasible, calculate 
   \begin{align}
       {\rm {dif}_{H}}=\frac{| T_h-T_{h(n)}|}{| T_h|}, ~~
       {\rm {dif}_{L}}=\frac{| T_l-T_{l(n)}|}{| T_l|}. \notag
   \end{align}
   If ${\rm {dif}_{H}} \le \epsilon_2$, ${\rm {dif}_{L}} \le \epsilon_2$ or $n_i$ is larger than a threshold, go to the next step.
   Or else set $T_{h(n)}:=T_{h}$, $T_{l(n)}:=T_{l}$, and $\boldsymbol{\eta_{(n)}}:=\boldsymbol{\eta}$, go to Step $2$.
   \item[(4)] If the problem in Step $2$ is feasible, set $t_{\min}:=t$; else set $t_{\rm max}:=t$.
   \item[(5)] Stop if $t_{\rm max}-t_{\rm min}< \epsilon_1$. Otherwise, initialize $n_i=0$, $T_{h(n)}$, $T_{l(n)}$, $\boldsymbol{\eta_{(n)}}$, and go to Step $2$.
\end{itemize}
\label{a_1}
\end{algorithm}
\setlength{\textfloatsep}{0.7cm}

\section{Numerical Results}~\label{sec:num}
In this section, numerical results are presented to examine the performance of the proposed hybrid OTFS/OFDM modulation system using the FZF and PZF precoding designs, as well as demonstrate the benefit of our power allocation frameworks.

\vspace{-1em}
\subsection{Large-scale Fading Model}
In our simulations, we consider a more practical large-scale fading system taking into account correlated shadowing~\cite{HienVS}. Note that this correlation may affect the system performance significantly.
We first assume that the users and BS are located over a $D \times D~{\rm km}^2$ space with uniform probability. Therefore, with the consideration of the path loss and shadow fading correlation model, the large-scale fading coefficient for the $k$-th user $\beta_{k}$ can be represented by
\begin{align}
    \beta_{k}={\rm PL}_k \times 10^{\frac{\sigma_{\rm sh} z_k}{10}},
    \label{LSfading}
\end{align}
where ${\rm PL}_k$ is the path loss coefficient, and $10^{\frac{\sigma_{\rm sh} z_k}{10}}$ models the shadowing effect with the standard deviation $\sigma_{\rm sh}$ and $z_k\sim \mathcal{N}(0, 1)$. We consider the three-slope path loss model in this paper~\cite{HienVS}. Specifically, the path loss exponent depends on the distance between the BS and the user $d_k$, and the path loss in dB can be represented as
\begin{align}
{\rm PL}_k\! =\!
 \begin{cases}
    -L-\!35\log_{10}(d_k), & {\rm if} \,\, d_k > d_1\\
    -L-\!15\log_{10}(d_1)-20\log_{10}(d_k), & {\rm if} \,\, d_1 \ge d_k > d_0\\
    -L-\!15\log_{10}(d_1)-20\log_{10}(d_0), & {\rm if} \,\, d_0\ge d_k,
 \end{cases}
\end{align}
with
\begin{align}
    L&\triangleq 46.3+33.9 \log_{10} (f) -13.82 \log_{10} (h_{\rm BS}) \notag \\
    &~~-(1.1 \log_{10} (f) -0.7) h_{\rm u} +(1.56 \log_{10}(f)-0.8).
\end{align}
Note that $f$ is the carrier frequency (in MHz), $h_{\rm BS}$ is the height of the BS antenna (in m), and $h_{\rm u}$ is the height of the user antenna (in m).

In practice, closely-located users may be surrounded by similar obstacles, and hence experience correlated shadowing. We consider a correlated shadowing effect for users with $d_k>d_1$, which can be denoted by~\cite{HienVS}
\begin{align}
    z_k= \sqrt{\delta} a + \sqrt{1-\delta} b_k,
\end{align}
where $\delta$, with $0\le \delta \le 1$, is a weighting parameter, and $a\sim \mathcal{N}(0, 1)$ and $b_k\sim \mathcal{N}(0, 1)$ are two random variables, modeling the shadowing from obstructing objects around the BS and $k$-th user, respectively. 
Specifically, we have 
\begin{align}
    \mathbb E \left\{  b_k b_{k'}\right\} =2^{- \frac{d(k,{k'})}{d_{\rm decorr}}},
\end{align}
where $d(k,{k'})$ is the geographical distance between the $k$-th and $k'$-th user, and $d_{\rm decorr}$ is the decorrelation distance. 

 \begin{table}
    \centering
        \caption{System Parameters for the Simulation }
    \begin{tabular}{|c|c|}
    \hline
        Parameter         & Value\\
          \hline
        Carrier frequency & $2$ GHz\\
          \hline
        Bandwidth & $20$ MHz\\
          \hline
        DL transmit power & $200$ mW\\
          \hline
        DL noise figure & $9$ dB\\
          \hline
        BS antenna height $h_{\rm BS}$ & $15$ m\\
           \hline
        User antenna height $h_{\rm u}$ & $1.65$ m\\
         \hline
        $\sigma_{\rm sh}$ & $8$ dB\\
         \hline
        $D$, $d_1$, $d_0$, $d_{\rm decorr}$ & $250$, $50$, $10$, $100$ m\\
         \hline
        Weighting parameter $\delta$ & 0.5 \\
        \hline
    \end{tabular}
    \label{tab:my_label}
\end{table}
\vspace{-1em}
\subsection{System Parameters}
Without loss of generality, we consider an OTFS transmission with $M=8$ and $N=8$. We set $K_h=3$, $K_l=3$ and $N_t=100$. We consider $P=3$ individual paths for each user with a uniform power delay profile. Similar to~\cite{Li2021performance,li2021cross}, the  delay $l_{k(i)}$ and Doppler indices $k_{k (i)}$ are generated with equal probability within the range of $[0, l_{max}]$ and $[-k_{max}, k_{max}]$, where the maximum delay index $l_{max}=3$ and the maximum Doppler index $k_{max}=3$ for LM-UEs, while $l_{max}=3$ and $k_{max}=5$ for HM-UEs, respectively. The CP length for OFDM users, which is set as $3$ in the paper, is decided by the maximum delay index.
To avoid the boundary effects and infinite simulation area, we assume that the simulation square area is wrapped around the edges with $8$ neighbors.
Moreover, the corresponding normalized transmit SNR $\rho$ can be calculated by dividing the DL transmit power by the noise power, where the noise power can be represented as
\begin{align}
    {\text {noise power}}={\text {bandwidth}}\times k_{\rm B} \times T_0 \times {\text {noise figure}}~{\text{(W)}}. \notag
\end{align}
Note that the Boltzmann constant $k_{\rm B}=1.381\times 10^{-23}$ (Joule per Kelvin), and the noise temperature $T_0=290$ (Kelvin). The other system parameters are set as shown in Table~\ref{tab:my_label}.

\subsection{Results and Discussions}
\begin{figure}
\vspace{-1em}
    \centering
    \includegraphics[width=0.9\linewidth]{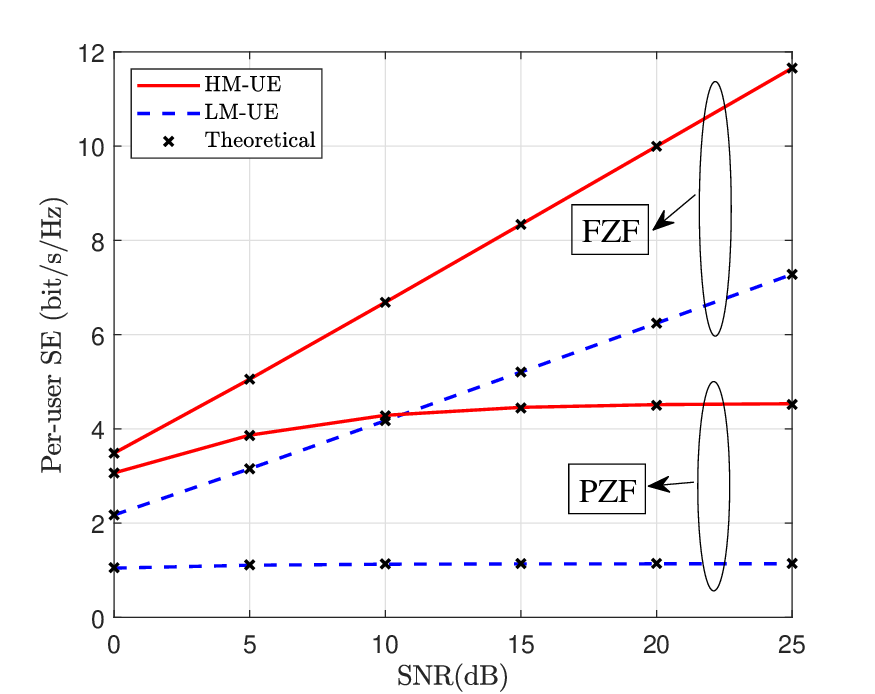}
    \caption{Theoretical and numerical per-user SE with $M=8$.}
    \label{fig:1}
\end{figure}
\begin{figure}
\centering
\includegraphics[width=0.9\linewidth]{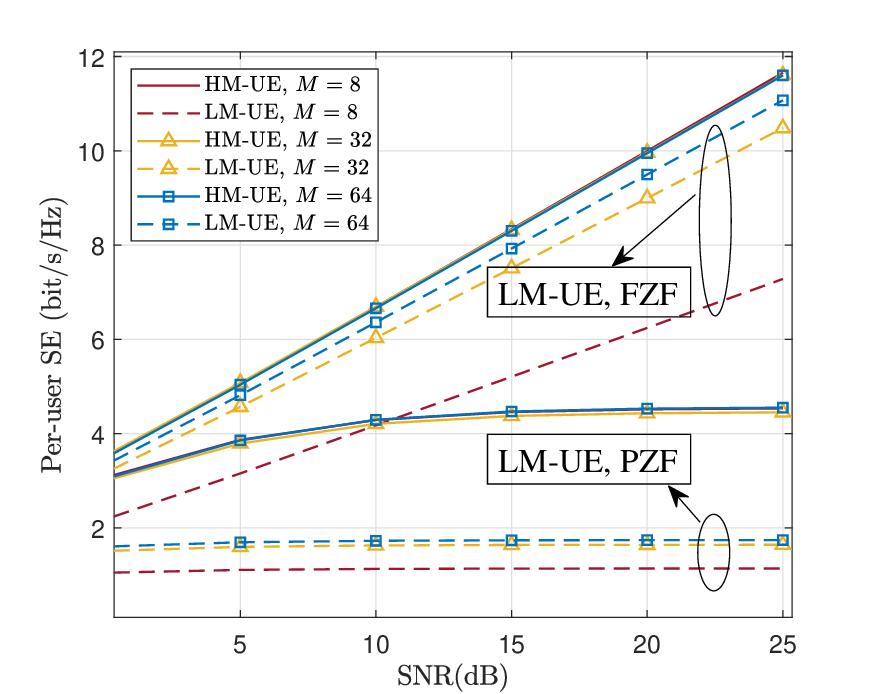}
\caption{Theoretical per-user SE with different $M $ values.}
\label{DifferentM}
\vspace{-2em}
\end{figure}
\begin{figure}[t]
    \centering
    \includegraphics[width=0.9\linewidth]{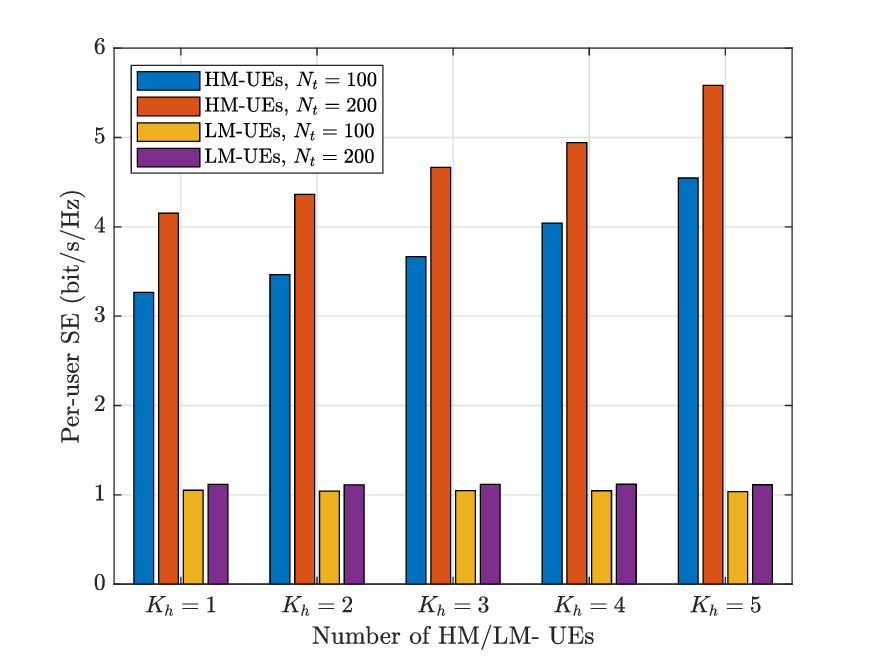}
    \vspace{0em}
    \caption{Per-user SEs with different numbers of users per each group ($K=6$, $K_l=K-K_h$).}
    \label{fig:7}
    \vspace{-0.9em}
\end{figure}
In Fig.~\ref{fig:1}, we compare the performance between FZF and PZF with equal power allocation (EPA) with $\beta_k=1$ and $\eta_k=\frac{1}{K}$ for each user. The simulation results verify the tightness of our derived closed-form SE approximation in Corollary $1$ and $2$. 
From Fig.~\ref{fig:1}, it is evident that FZF precoding offers performance enhancements over PZF precoding for both HM- and LM-UEs. Additionally, HM-UEs consistently demonstrate superior performance compared to LM-UEs across both precoding schemes. This discrepancy arises primarily because, although FZF effectively cancels out all interference for all users, LM-UEs experience a lower SE due to the CP insertion inherent in OFDM modulation. Furthermore, under PZF, LM-UEs suffer from increased interference compared to HM-UEs, in addition to variations in the CP overhead levels.

In Fig.~5, we show the effect of different frame sizes on the per-user SEs with the proposed precoding schemes. 
By considering the per-user SE and the MMSE-SIC detection, the different frame sizes only affect the CP overhead level, which is $\frac{L_{CP} (N+1)}{MN+L_{CP}}$ for the LM-UEs, and $\frac{L_{CP}}{MN+L_{CP}}$ for the HM-UEs. Therefore, in Fig.~5, to better understand the effect of the overhead, we consider systems with the same $N=8$ and different $M$ values to ensure a fixed CP length for both HM-UEs and LM-UEs. From the simulation results, we can see that with different frame lengths, the per-user SE has the same trend for different users. Specifically, for HM-UEs, the per-user SE illustrates a similar performance for different frame sizes, while with larger $M$, the LM-UEs have a linear improvement in performance. 
This is because, with the same CP length, the overhead level decreases with larger $M$ values. Note that the decrease of the overhead levels becomes marginal when the CP length is relatively small compared to the frame size. 
Therefore, due to the reduced CP structure for OTFS, the performance is similar for different frame sizes. 
For OFDM, we can see a noticeable performance improvement from $M=8$ to $M=32$. However, the performance improvement becomes marginal when we further increase the $M$ value. 
Based on these observations, and without loss of generality, we mainly consider $M=N=8$ for our simulations in the following parts for simplicity.

To shed light on the trade-off between the computational complexity and performance, we then consider different numbers of users in each group without the large-scale fading and power allocation, as shown in Fig.~\ref{fig:7}. As the PZF precoding complexity is dependent on the number of HM-UEs, the numerical results illustrate that with more HM-UEs, higher SE can be achieved by the HM-UEs at the cost of high complexity. However, the performance for LM-UEs remains the same, as they are suffering from both inter- and intra-group interference. Moreover, we can notice that, with a larger number of $N_t$, a better performance can be achieved by all users.

\begin{figure}
\centering
\begin{subfigure}[a]{0.5\textwidth}
\centering
\includegraphics[width=0.9\linewidth]{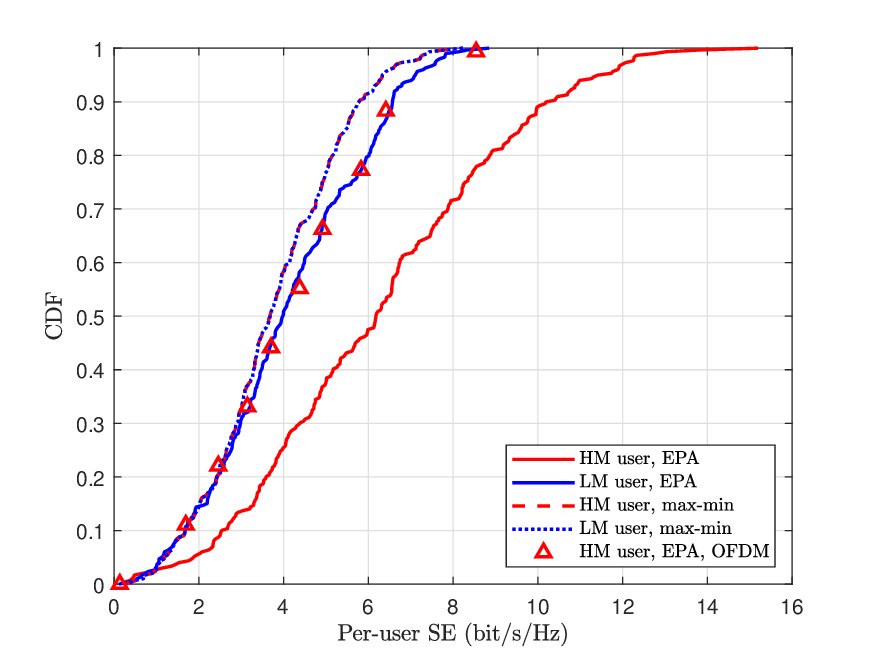}
\caption{ FZF precoding.}
\label{fig:2}
\end{subfigure}
\begin{subfigure}[a]{0.5\textwidth}
\centering
\includegraphics[width=0.9\linewidth]{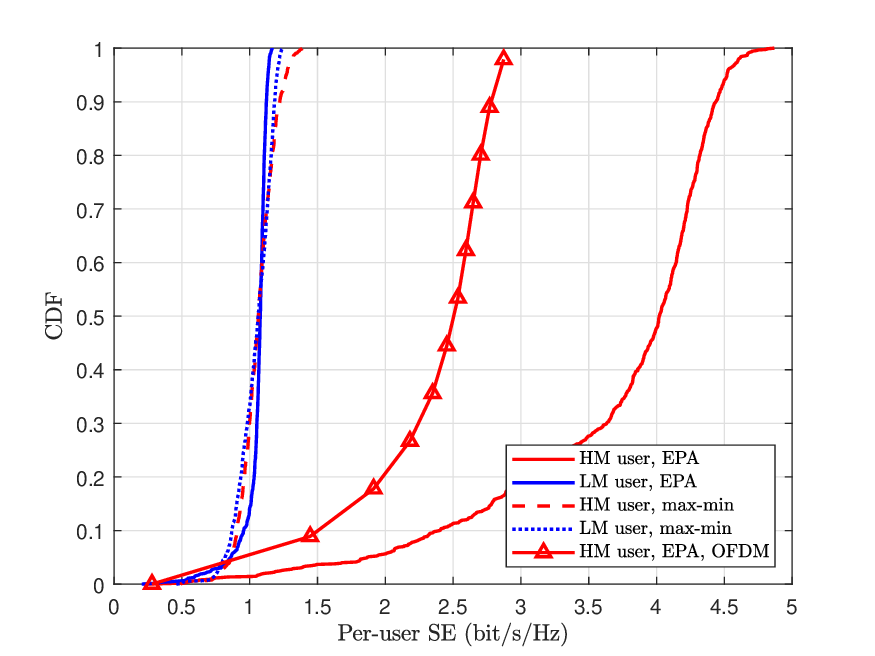}
\caption{PZF precoding.}
\label{fig:3}
\end{subfigure}
\caption{Per-user SE with max-min power allocation.}
\vspace{-0.8em}
\end{figure}
To further demonstrate the fairness of the performance for all users, we then show the simulation results for FZF and PZF with max-min power allocation in Fig.~\ref{fig:2} and Fig.~\ref{fig:3}, respectively. 
In this paper, we set $t_{min}$ as $0$, while the specific value for $t_{max}$ depends on the network setup and parameters. Therefore, we approximate the value by using a multiple of the SE with equal power allocation.
Moreover, the performance of the HM-UE with OFDM modulation is given in the simulation results as a benchmark. We can clearly see the performance enhancement provided by using OTFS over OFDM for HM-UEs. The similar performance for HM-UEs with OFDM and LM-UEs with OFDM under the FZF precoding verifies our previous discussion on the performance loss for LM-UEs with OFDM due to the CP overhead. From Fig.~\ref{fig:2} and Fig.~\ref{fig:3}, we can observe that the HM-UEs have better performance than the LM-UEs with equal power for all users. After applying the max-min power allocation, HM and LM-UEs end up with the same system performance that is similar to the performance of LM-UEs with uniform power allocation.  This is due to the fact that by achieving fairness between all users, we compensate the SE of all users to eliminate the performance gap between different groups of users. 

\begin{figure}[t]
    \centering
    \includegraphics[width=0.9\linewidth]{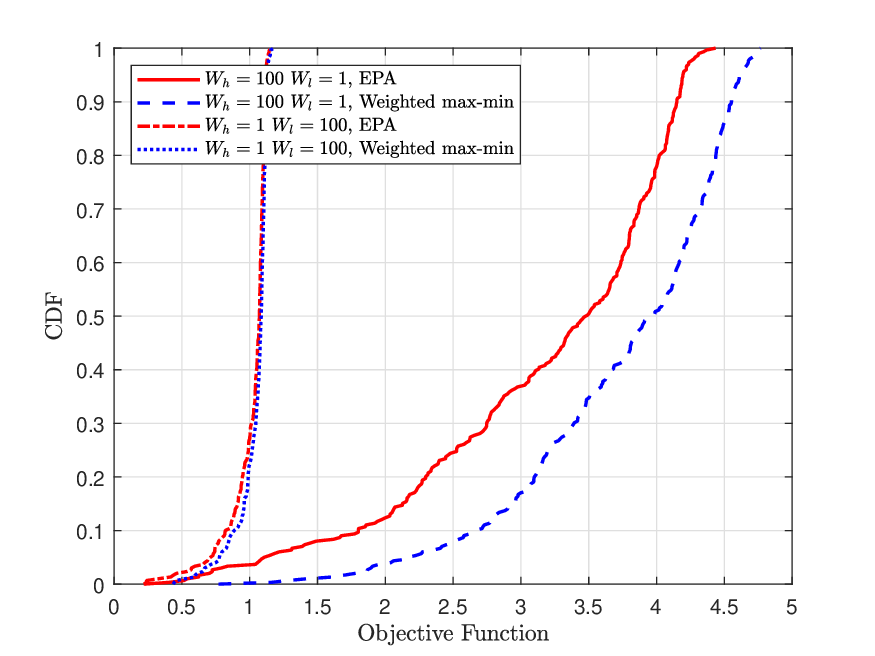}
    \caption{The value of the objective function in~\eqref{PC_1} with and without optimization.}
    \label{fig:4}
    \vspace{-0.5em}
\end{figure}
\begin{figure}[t]
\begin{subfigure}[a]{0.5\textwidth}
\includegraphics[width=0.9\linewidth]{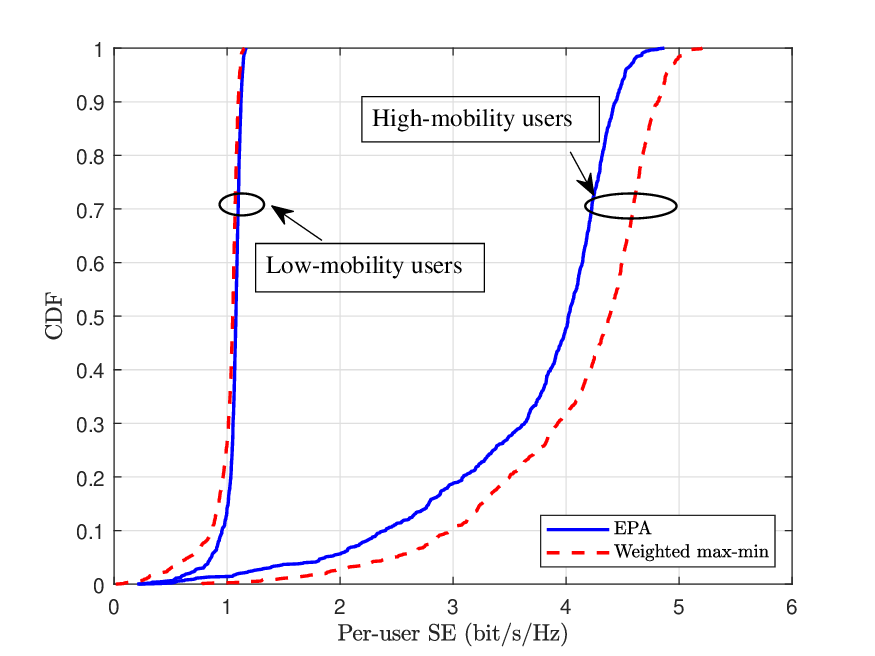}
\caption{$w_h=100$, $w_l=1$}
\label{fig:5}
\end{subfigure}
\begin{subfigure}[a]{0.5\textwidth}
\includegraphics[width=0.9\linewidth]{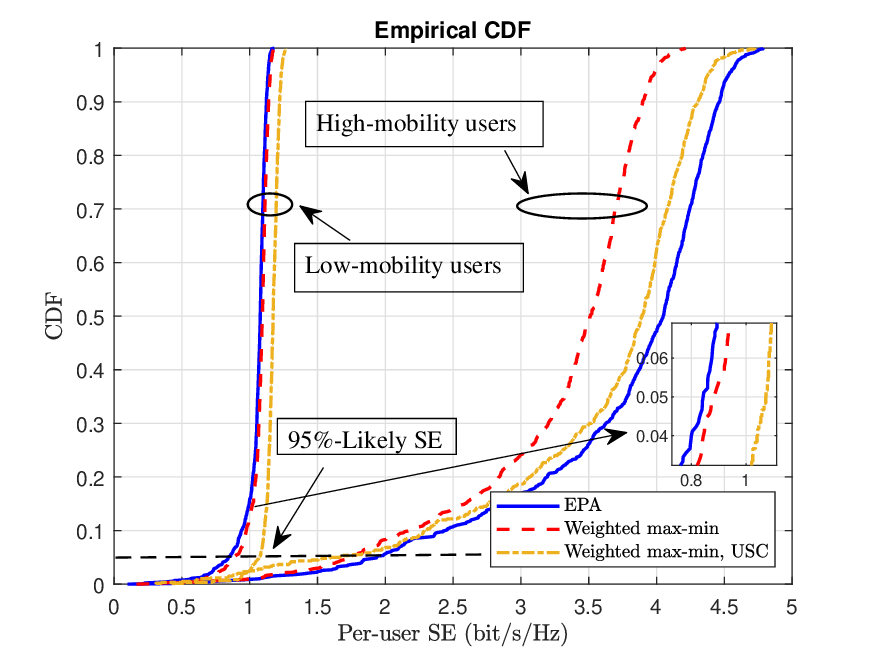}
\caption{$w_h=1$, $w_l=100$}
\label{fig:6}
\end{subfigure}
\caption{Per-user SE for HM- and LM-UEs with weighted max-min power allocation.}
\vspace{-0.5em}
\end{figure}

    
With the substantial performance differences between HM and LM-UEs with PZF precoding caused by the different interference levels, promoting fairness among all users compensates too much performance of the HM-UEs. 
Therefore, we then show the simulation results for weighted max-min power allocation in Fig.~\ref{fig:4},~\ref{fig:5},~\ref{fig:6}, where fairness is considered among HM-UEs and LM-UEs, respectively. In Fig.~\ref{fig:4}, we show the objective function value before and after the weighted max-min power allocation with two different sets of weighting coefficients. The simulation results show the improvement after the weighted max-min power allocation, indicating the efficiency of Algorithm $1$.

In Fig.~\ref{fig:5}, we compare the performance with and without the weighted max-min power allocation, with $w_h=100$ and $w_l=1$ specifically. With the priority given to the HM-UEs, we can notice the performance improvement after the power allocation for HM-UEs. In parallel, a decrease in the performance of the LM-UEs can be observed. 
On the other hand, we give priority to the LM-UEs by setting $w_h=1$ and $w_l=100$ in Fig.~\ref{fig:6}. From the simulation results, we see a non-negligible performance improvement for the LM-UEs. 
This suggests that, with the weighted max-min power allocation, fairness can be achieved by the users in the groups of HM and LM, respectively. Additionally, by changing the weighting coefficients, priority can be given to one of the considered groups, resulting in a performance improvement for the prioritized group. 
Since the performance improvement is minor for LM-UEs as the prioritized group in Fig.~\ref{fig:6}, we then consider the weighted max-min method with USC. 
By scheduling the LM-UE with the lowest theoretical SE based on the statistical CSI, around $20\%$ performance improvement in the $95\%$-likely SE can be achieved for LM-UEs.

\section{Conclusion}~\label{sec:conc}
We investigated a DL massive MIMO system following a hybrid OTFS/OFDM transmission protocol. With the user grouping based on the users' mobility profile, two different precoding schemes were considered. The performance of the system was investigated based on the MMSE-SIC detection in terms of SE. We showed that the FZF eliminates all the interference at the cost of high complexity. For PZF, the inter-group interference for the HM-UEs can be eliminated with a reduced complexity. We also observed that the LM-UEs are affected by both inter- and intra-group interference. To further enhance the fairness among users, we applied the max-min power allocation for all users with FZF and PZF, respectively. Due to the significant performance gap between HM with PZF, a weighted max-min power allocation scheme was also considered. 
Our simulation results validated our theoretical analysis and illuminated some practical guidelines for OTFS/OFDM-massive MIMO systems with different complexity and performance levels. 
As part of future work, the estimation methods for the angle of departure/arrival and the effects of the corresponding estimated angular error could be investigated.
\appendices
\section{Proof of Proposition 1}
\label{APX:theorem:FZF:LM}
By invoking~\eqref{eq:SEk} and~\eqref{FZF}, for the $k_h$-th HM-UE, with $k_h'\in \Kh$, we have
\begin{align}
\D_{k_h k_h'}
&=\sqrt{\rho\eta_{k_h'}} ({\bf F}_{N} \otimes {\bf I}_M)   \mH _{k_h}\qW_{k_h'}^\FZF ({\bf F}_{N}^{\rm H} \otimes {\bf I}_M)	\notag\\
&=\alpha_{\FZF}\sqrt{\rho\eta_{k_h'}} ({\bf F}_{N} \otimes {\bf I}_M) \mH _{k_h} ({\mH}^{\rm FZF})^{\rm H} \notag\\
&~~\times \left (  {\mH}^{\rm FZF} ({\mH}^{\rm FZF})^{\rm H} \right )^{-1} {\bf B}_{k_h'} ({\bf F}_{N}^{\rm H} \otimes {\bf I}_M) \notag \\
&= \alpha_{\FZF}\sqrt{\rho\eta_{k_h'}}  ({\bf F}_{N} \otimes {\bf I}_M) {\bf B}_{k_h}^{\rm H}  {\mH}^{\rm FZF} ({\mH}^{\rm FZF})^{\rm H}   \notag \\
&~~\times \left (  {\mH}^{\rm FZF} ({\mH}^{\rm FZF})^{\rm H} \right )^{-1} {\bf B}_{k_h'} ({\bf F}_{N}^{\rm H} \otimes {\bf I}_M) \notag \\
&\stackrel{(a)}{=}
 \begin{cases}
    {\bf 0}_{MN}, & k_h'\neq k_h\\
\alpha_{\FZF}\sqrt{\rho\eta_{k_h}}{\bf I}_{MN}, & k_h'= k_h
 \end{cases},
\label{FZF_HW_H}
\end{align}
where $(a)$ in \eqref{FZF_HW_H} is due to the structure of ${\bf B}_{k_h}^{\rm H}$ and ${\bf B}_{k_h'}$.
Similarly, for $k_l'\in \Kl$, we have
\begin{align}
  \D_{k_h k_l'}
  &=\sqrt{\rho\eta_{k_l'}} ({\bf F}_{N} \otimes {\bf I}_M) \mH _{k_h}\qW_{k_l'}^\FZF \left (  {\bf I}_{N}\otimes{{\bf A}_{\rm CP}{\bf{F}}^{\rm H}_{L_d}}  \right ) \notag\\
  &=\alpha_{\FZF} \sqrt{\rho\eta_{k_l'}} ({\bf F}_{N} \otimes {\bf I}_M)  {\bf B}_{k_h}^{\rm H}  {\mH}^{\rm FZF} ({\mH}^{\rm FZF})^{\rm H}   \notag \\
  &~~\left (  {\mH}^{\rm FZF} ({\mH}^{\rm FZF})^{\rm H} \right )^{-1} {\bf B}_{k_l'} \left (  {\bf I}_{N}\otimes{{\bf A}_{\rm CP}{\bf{F}}^{\rm H}_{L_d}}  \right )   \notag \\
  &={\bf 0}_{MN\times L_d N}.
\end{align}
Therefore, for $k'\in\{ 1,\ldots, K \}$ and $k'\neq k_h$, we have 
\begin{align}
\D_{k_h k'} \D_{k_h k'}^{\rm H}={\bf 0}_{MN}.
\label{FZF_HW_H_1}
\end{align}
Based on~\eqref{FZF_HW_H} and~\eqref{FZF_HW_H_1}, we have
\begin{align}~\label{eq:Dkhkh}
	\BD_{k_h k_h}=\alpha_{\FZF} \sqrt{\rho\eta_{k_h}} {\bf I}_{MN},
\end{align}
and,
\begin{align}~\label{eq:SiFZF}
{\bf \Psi}_{k_h} 
&= {\bf I}_{MN} + \mathbb E \big\{ \D_{k k} \D_{k k}^{\rm H}  \big\} +(K-1){\bf 0}_{MN}- \BD_{k k} \BDH_{k k} \notag \\
&= {\bf I}_{MN}.
\end{align}
Hence, by substituting~\eqref{eq:Dkhkh} and~\eqref{eq:SiFZF} into~\eqref{eq:SEk}, the SE for $k_h$-th HM-UE can be obtained as~\eqref{SE_HFZF}. 

Similarly, for the $k_l$-th LM-UE, with $k_l'\in \Kl$, we have 
\begin{align}
\D_{k_l k_l'}&=\sqrt{\rho \eta_{k_l'}}   \left(  {\bf I}_{N}\otimes{{\bf{F}}_{L_{d}}}{{\bf R}_{\rm CP}}  \right)  \Htd _{k_l} {\bf W}_{k_l'} \left (  {\bf I}_{N}\otimes{{\bf A}_{\rm CP}{\bf{F}}^{\rm H}_{L_d}}  \right ) \notag\\
&= \alpha_{\FZF} \sqrt{\rho\eta_{k_l'}}   \left(  {\bf I}_{N}\otimes{{\bf{F}}_{L_{d}}}{{\bf R}_{\rm CP}}  \right)  \Htd _{k_l} ({\mH}^{\rm FZF})^{\rm H} \notag \\
&~~\times\left (  {\mH}^{\rm FZF} ({\mH}^{\rm FZF})^{\rm H} \right )^{-1} {\bf B}_{k_l} \left (  {\bf I}_{N}\otimes{{\bf A}_{\rm CP}{\bf{F}}^{\rm H}_{L_d}}  \right ) \notag \\
&=\begin{cases}
	{\bf 0}_{L_dN}, & k_l'\neq k_l\\
		\alpha_{\FZF} \sqrt{\rho\eta_{k_l}}{\bf I}_{L_dN}, & k_l'= k_l
  \end{cases}.
\label{FZF_HW_L}
\end{align}
For $k_h\in\Kh$,
\vspace{-0.4em}
\begin{align}
  \D_{k_l k_h}
  &=\sqrt{\rho\eta_{k_h}} \left(  {\bf I}_{N}\otimes{{\bf{F}}_{L_{d}}}{{\bf R}_{\rm CP}}  \right)  \Htd _{k_l}\qW_{k_h}^\FZF \left( {{{\bf{F}}_N^{\rm H}} \otimes {{\bf{I}}_M}} \right) \notag\\
  &={\bf 0}_{L_d N\times M N}.
\end{align}
Therefore, for $k'\in\{ 1,\ldots, K \}$ and $k'\neq k_l$, we have 
\begin{align}
\D_{k_l k'} \D_{k_l k'}^{\rm H}={\bf 0}_{L_dN}.
\end{align}
Then, the SE for the $k_l$-th LM-UE can be obtained as~\eqref{SE_LFZF}.

\section{Proof of Proposition 2}
\label{Ptheorem:PZFMRT}
Focusing on the $k_h$-th HM-UE with $k_h'\in \Kh$, similar as in~\eqref{FZF_HW_H} and~\eqref{FZF_HW_L}, we have
 \begin{align}
\D_{k_h k_h'}
&= \alpha^{\PZF}_{k'_h} \sqrt{\rho\eta_{k_h'}}  ({\bf{F}}_N \otimes {\bf{I}}_M) \Htd _{k_h} (\mH^\PZF)^{\rm H} \notag \\
&~~ \times \left (  \mH^\PZF (\mH^\PZF)^{\rm H} \right )^{-1} \left ( {\qb_{K_h'}^{(k_h)}} \otimes \qI_{MN} \right) \left( {{{\bf{F}}_N^{\rm H}} \otimes {{\bf{I}}_M}} \right) \notag \\
&=
\begin{cases}
{\bf 0}_{MN}, & k_h'\neq k_h\\
\alpha^{\PZF}_{k_h}\sqrt{\rho\eta_{k_h}}{\bf I}_{MN}, & k_h'= k_h
\end{cases}.
\end{align} 
Therefore, we have
\begin{align}
\mathbb E \Big\{ \D_{k_h k_h} \D_{k_h k_h}^{\rm H}  \Big\}
=(\alpha^{\PZF}_{k_h})^2 {\rho\eta_{k_h}} {\bf I}_{MN}.
\end{align}
Moreover, for the inter-group interference from user $k_l'$, with $k_l\in \Kl$, 
\begin{align}
    \D_{k_h k_l'}=\sqrt{\rho\eta_{k_l'}}  ({\bf{F}}_N \otimes {\bf{I}}_M)  \Htd _{k_h}  {\bf W}_{k_l'}^{\MRT}  \left (  {\bf I}_{N}\otimes{{\bf A}_{\rm CP}{\bf{F}}^{\rm H}_{L_d}}  \right ).
\end{align}
Then, we have 
\begin{align}
\label{Dkkp_1}
&\mathbb E \left\{ \D_{k_h k_l'} \D_{k_h k_l'}^{\rm H}  \right\} \notag\\
&~~=(\alpha^\MRT_{k'_l}) ^2\rho\eta_{k_l'} ({\bf{F}}_N \otimes {\bf{I}}_M) \mathbb E \Big\{\Htd _{k_h}   (\Htd _{k_l'})^{\rm H}   \notag \\
&~~~~ \times \left (  {\bf I}_{N}\otimes{{\bf A}_{\rm CP}{\bf{F}}^{\rm H}_{L_d}}{\bf{F}}_{L_d} {\bf A}_{\rm CP}^{\rm H} \right )   \Htd _{k_l'} (\Htd _{k_h})^{\rm H}\Big\} ({\bf{F}}_N^{\rm H} \otimes {\bf{I}}_M) \notag\\
&~~=(\alpha^\MRT_{k'_l}) ^2\rho\eta_{k_l'} ({\bf{F}}_N \otimes {\bf{I}}_M) \mathbb E \Big\{\Htd _{k_h}   (\Htd _{k_l'})^{\rm H}   \notag \\
&~~~~ \times \left (  {\bf I}_{N}\otimes{\bf A}_{\rm CP} {\bf A}_{\rm CP}^{\rm H} \right )   \Htd _{k_l'} (\Htd _{k_h})^{\rm H}\Big\} ({\bf{F}}_N^{\rm H} \otimes {\bf{I}}_M).
\end{align} 
Based on~\eqref{21c}, we have
\vspace{-0.9em}
\begin{align}
{\bf \Psi}_{k_h} 
= {\bf I}_{MN} + \sum^{K_l}_{k_l'=1} \mathbb E \big\{ \D_{k_h k'_l} \D_{k_h k'_l}^{\rm H}  \big\} ,
\end{align}
and~\eqref{SE_PZF_1} can then be obtained.

\section{Proof of Proposition 3}
\label{APX:4}
For the $k_l$-th LM-UE, by invoking~\eqref{W_MRT},  we have
\begin{align}
&\D_{k_l k_l} 
=   \sqrt{\rho \eta_{k_l}}    \left(  {\bf I}_{N}\otimes{{\bf{F}}_{L_{d}}}{{\bf R}_{\rm CP}}  \right)  \Htd _{k_l} {\bf W}_{k_l} \left (  {\bf I}_{N}\otimes{{\bf A}_{\rm CP}{\bf{F}}^{\rm H}_{L_d}}  \right )  \notag\\
&\hspace{3em}= \alpha^{\MRT}_{k_l} \sqrt{\rho \eta_{k_l}}    \left(  {\bf I}_{N}\otimes{{\bf{F}}_{L_{d}}}{{\bf R}_{\rm CP}}  \right)  
\nonumber\\
&\hspace{4em}\times\Htd _{k_l} \big(\Htd _{k_l}\big)^{\rm H}\!\! \left (  {\bf I}_{N}\otimes{{\bf A}_{\rm CP}{\bf{F}}^{\rm H}_{L_d}}  \right )\!.
\end{align}
Similar to~\eqref{a_n_mrt}, we have
\begin{align}
    &\mathbb{E}\Big\{ \Htd _{k_l} \big(\Htd _{k_l}\big)^{\rm H}\Big\} \notag\\
    &= \beta _{k_l} \mathbb{E}\Big\{ \Big ( \sum_{i=1}^{P} {\bm \theta} _{k_l(i)} \otimes {\bf H}^{\rm TD} _{k_l(i)} \Big ) \Big ( \sum_{j=1}^{P} {\bm \theta} _{k_l(j)} \otimes {\bf H}^{\rm TD} _{k_l(j)} \Big )^{\rm H} \Big\} \notag\\
    &=\beta _{k_l} \sum\nolimits_{i=1}^{P} \mathbb E \left\{ {\bm \theta} _{k_l(i)} {\bm \theta} _{k_l(i)}^{\rm H} \right\} \otimes \mathbb E \left\{ {h_{k_l (i)}}{h_{k_l (i)}^{\rm H}}  {\bf I}_{MN} \right\} \notag\\
    &=\beta _{k_l} N_t  {\bf I}_{MN}.
\end{align}
Therefore, we have
\begin{align}
\BD_{k_l k_l} 
&=  \alpha^{\MRT}_{k_l} \sqrt{\rho \eta_{k_l}} \left(  {\bf I}_{N}\otimes{{\bf{F}}_{L_{d}}}{{\bf R}_{\rm CP}}  \right) \mathbb{E}\left\{ \Htd _{k_l} (\Htd _{k_l})^{\rm H}\right\}    \notag \\
&\hspace{3em}\times \left (  {\bf I}_{N}\otimes{{\bf A}_{\rm CP}{\bf{F}}^{\rm H}_{L_d}}  \right ) \notag\\
&=\alpha^{\MRT}_{k_l} \sqrt{\rho \eta_{k_l}} \beta _{k_l} N_t  {\bf I}_{L_dN} .
\end{align} 
To this end, after computing ${\bf \Psi}_k$ according to~\eqref{21c} and then plugging the result into~\eqref{eq:SEk} we arrive at~\eqref{SE_PZF_2}.

\section{}
\label{APX:3}
For the first part, recall that $h_{k_l(i)} \sim \mathcal{CN}(0, \frac{1}{P})$, and $\mathbb E \left \{ \{\Re (h_{k_l(i)})\}^4\right \}=\mathbb E \left \{ \{\Im (h_{k_l(i)})\}^4\right \}= \frac{3}{4P^2} $. Therefore, we have
\begin{align}
&\beta _{k_l}^2 \sum_{i=1}^{P}  \mathbb E \left\{ {\bf{H}}^{\rm{TD}}_{k_l (i)}  ({\bf{H}}^{\rm{TD}}_{k_l (i)})^{\rm H}  {\bf{H}}^{\rm{TD}}_{k_l (i)}  ({\bf{H}}^{\rm{TD}}_{k_l (i)})^{\rm H}\right\}  \notag\\
&~~=\beta _{k_l}^2 \!\!\sum_{i=1}^{P} \!\mathbb E\! \left\{ | {\bm \theta} _{k_l(i)} {\bm \theta} _{k_l(i)}^{\rm H}| ^2\right\} \! \mathbb E \!\left\{  h_{k_l(i)} h_{k_l(i)}^* h_{k_l(i)} h_{k_l(i)}^*   \right\} \!{\bf I}_{MN}\notag\\
&~~=\beta _{k_l}^2  N_t^2 \frac{2}{P} {\bf I}_{MN}.
\end{align}
For the second part, similar to~\eqref{PZF_H_2_1}, we have 
\begin{align}
&\beta _{k_l}^2 \sum_{i=1}^{P} \sum_{j=1, j\neq i}^{P} \mathbb E \left\{ {\bf{H}}^{\rm{TD}}_{k_l (i)}  ({\bf{H}}^{\rm{TD}}_{k_l (j)})^{\rm H}  {\bf{H}}^{\rm{TD}}_{k_l (j)}  ({\bf{H}}^{\rm{TD}}_{k_l (i)})^{\rm H}\right\} \notag\\
&~~= \beta _{k_l}^2 (P^2-P) N_t \frac{1}{P^2}  {\bf I}_{MN} \notag\\
&~~= \beta _{k_l}^2 \frac{P-1}{P} N_t   {\bf I}_{MN} .
\end{align}
At last, we have
\begin{align}
&\beta _{k_l}^2 \sum_{i=1}^{P} \sum_{j=1, j\neq i}^{P} \mathbb E \left\{ {\bf{H}}^{\rm{TD}}_{k_l (i)}  ({\bf{H}}^{\rm{TD}}_{k_l (i)})^{\rm H}  {\bf{H}}^{\rm{TD}}_{k_l (j)}  ({\bf{H}}^{\rm{TD}}_{k_l (j)})^{\rm H}\right\} \notag \\
&=\beta _{k_l}^2 \sum_{i=1}^{P} \sum_{j=1, j\neq i}^{P} \!\mathbb E \left\{ {\bm \theta} _{k_l(i)} {\bm \theta} _{k_l(i)}^{\rm H} {\bm \theta} _{k_l(j)} {\bm \theta} ^{\rm H}_{k_l(j)}\right\} \notag \\ 
&~~~\times\mathbb E \left\{ h_{k_l(i)} h_{k_l(i)}^* h_{k_l(j)} h_{k_l(j)}^*  \right\} {\bf I}_{MN} \notag\\
&= \beta _{k_l}^2 \frac{P-1}{P} N_t^2   {\bf I}_{MN} .
\end{align}

\bibliographystyle{IEEEtran}
\bibliography{main}

\end{document}